\documentstyle[12pt]{article}
\begin{document}
\noindent
\hspace*{13.55cm}
DO-TH 97-13\\
\noindent
\hspace*{13.3cm}
SNUTP 97-078\\
\noindent
\hspace*{13.5cm}
YUMS 97-015\\
\noindent
\hspace*{13.15cm}
hep-ph/9706323\\

%\vspace{1.8cm}

\centerline{\Large \bf {One-loop renormalization group equations of}}
\centerline{\Large \bf
 {the general framework with two Higgs doublets}}

\vspace{0.8cm}

\begin{center}
{{\bf G.~Cveti\v c} \\ 
{\it Inst.~f\"ur Physik, Universit\"at Dortmund, 44221 Dortmund, Germany}}

\vspace{0.2cm}
{{\bf S.S.~Hwang} ~~~and~~~ {\bf C.S.~Kim}\\
{\it Department of Physics, Yonsei University, Seoul 120-749, Korea}}

\vspace{0.5cm}

(\today)
\end{center}

{\centerline {\bf {Abstract}}}

\vspace{0.2cm}

We derive one-loop renormalization group equations (RGE's) for
Yukawa coupling parameters of quarks and for the vacuum expectation
values of the Higgs doublets in a general framework 
of the Standard Model with two Higgs doublets (2HDM ``type III''). 
In the model, the neutral-Higgs-mediated
flavor-changing neutral currents
are allowed but are assumed to be reasonably suppressed 
at low energies. The popular ``type II'' and ``type I''
models are just special cases of this framework. We also present a 
numerical example for the RGE flow of Yukawa coupling parameters
and masses of quarks. Detailed investigation, through RGE's, of
connection between the low energy and the high
energy structure of these 2HDM's would give us some 
additional insights into the viability of such frameworks, 
{\it i.e.}, it would tell us for which choices of phenomenologically
acceptable values of low energy parameters
such frameworks can be regarded as reasonably natural
(-- without excessive changes in the parameter structure
as energy of probes increases over a wide range).
Furthermore, such analyses
could provide us with possible signals
of new physics at high energies.\\

\noindent PACS: 11.10.Hi; 12.15.Ff; 12.15.Mm; 12.38.Bx; 12.60.Fr\\

\vspace{0.8cm}

\small\normalsize

\section{Introduction}

Low energy experiments show that flavor-changing neutral currents (FCNC's) are
very suppressed in nature. For example,
\small
\begin{displaymath}
Br(K^0_L \to \mu^+\mu^-) \simeq (7.4 \pm 0.4) \cdot 10^{-9} \ , \qquad 
|m_{B^0_H}-m_{B^0_L} | \simeq (3.36 \pm 0.40) \cdot 10^{-10} \mbox{MeV} \ , 
\end{displaymath}
\begin{displaymath}
 |m_{K_L}-m_{K_S}| \simeq (3.510 \pm 0.018) \cdot 10^{-12} \mbox{MeV} \ , \ 
 |m_{D^0_1}-m_{D^0_2}| < 1.32 \cdot 10^{-10} \mbox{MeV} \ , \
\end{displaymath}
\begin{displaymath}
 Br(b \to s \gamma) = (2.32 \pm 0.67) \cdot 10^{-4} \ \mbox{{\it etc.}}
\end{displaymath}
\normalsize
The various alternative models of electroweak interactions --
extensions of the minimal Standard Model (MSM) -- must take into account
this FCNC suppression. The most conservative extensions of the MSM
are apparently the models with two Higgs doublets (2HDM's).
The conditions for the one-loop FCNC suppression of contributions
coming from gauge boson loops, {\it i.e.}, the allowed representations
of fermions, have been investigated some time
ago~\cite{PaschosGlashowWeinberg}. 
In addition to the MSM (one Higgs doublet model),  
Glashow and Weinberg~\cite{PaschosGlashowWeinberg} proposed 
for the Higgs sector 
the ``type I'' and ``type II'' 2HDM's. They proposed them as 
models which,  in a ``natural'' way, have the zero value for the
flavor-changing renormalized ({\it i.e.}, low energy)
Yukawa couplings in the neutral sector 
(called from now on: FCNC renormalized Yukawa couplings).
These two types of the 2HDM's have
been widely discussed in the literature.

Later on, extensions with more than one Higgs doublet, other than
the 2HDM(I) and (II), have been proposed and investigated
-- the general ``type III'' 2HDM's 
\cite{Lee}-\cite{BarShalomEilamSoniWudka}.
This is the framework in which the renormalized
({\it i.e.}, low energy) FCNC Yukawa coupling\footnote{
A more precise expression would be ``neutral flavor-changing
scalar (Yukawa) coupling,'' since these couplings
have no four-vector current structure involving $\gamma^{\mu}$.} 
parameters are in general nonzero, but must be sufficiently suppressed.
By ``sufficiently'' we mean 
that the measured low energy flavor-changing processes,
including the $\triangle F =2$
processes, $K^0$-${\overline {K}}^0$, $B^0$-${\overline {B}}^0$ and
$D^0$-${\overline {D}}^0$, do not get enhanced by tree level 
neutral Higgs exchange diagrams beyond the
bounds given by low energy experiments. However, 
it may well be that the suppression of FCNC's is
not relevant for the top quark, {\it i.e.}, we may have
appreciable renormalized
coupling of $t$ and $c$ to neutral scalars.
This is so because, up to date, no stringent experimental bounds for 
these direct FCNC couplings of the heavy top quarks exist.

The general 2HDM(III) framework has been
first mentioned already in 1973 by T.D.~Lee \cite{Lee},
for the hadronic sector and with the emphasis on
the CP-violating phenomena originating from the nonzero
relative phase of the two vacuum expectation values (VEV's). 
Later on, the model has been
investigated by several authors 
\cite{Sikivie}-\cite{LiuWolfenstein}, \cite{WuWolfenstein},
who mainly investigated the bounds on the low energy
({\it i.e.}, renormalized) parameters in the scalar mass and
in the Yukawa parameter sector -- the bounds resulting from
available phenomenological data on CP-violating phenomena,
such as ${\varepsilon}$ and ${\varepsilon}^{\prime}$
parameters of the kaon physics, and the neutron
electric dipole moment.

In 2HSM(III), specific ans\"atze for the FCNC Yukawa coupling 
parameters\footnote{
Implicitly, all these coupling parameters are the
renormalized (or: nearly renormalized) parameters, {\it i.e.},
they are parameters at the low evolution energies
since the authors investigated their tree level contributions
to (low energy) FCNC phenomena for which 
phenomenological data are available.}
have been proposed by Cheng, Sher and Yuan (CSY)
\cite{ChengSher}, \cite{SherYuan}, and by Antaramian, Hall and
Ra\v sin (AHR) \cite{AntaramianHallRasin}. These two
groups of authors also investigated bounds on
the masses of scalars and on FCNC Yukawa coupling parameters
arising from available phenomenological data
of FCNC phenomena, such as the $K$-${\overline {K}}$
and $B$-${\overline {B}}$ mass differences and rare $B$ decays.
The basic messages of their works are the following:
\begin{itemize}
\item
their ans\"atze are reasonably natural (or, more cautiously:
not ``unnatural'') from the
aspect of actual hierarchy of fermionic masses,
since they are motivated to a large degree
by this mass hierarchy.
\item
their ans\"atze allow the masses of neutral
scalars to be as low as $\sim 10^2$ GeV
while still not violating available (low energy) data
of FCNC phenomena.
\end{itemize}
CSY ansatz is given explicitly in Section 2.
Later on, several authors further investigated 
implications of the CSY, AHR and/or related 
ans\"atze, and of specific assumed ranges of 
masses of scalars, for the FCNC phenomena measured presently
or to be measured in various possible future colliders,
such as: (tree level effects in)
the decays $B \to {\mu}^+{\mu}^-$ \cite{Savage};
one-loop processes $t \to c \gamma$ and $t \to c Z$ \cite{LukeSavage};
two-loop effects in $\mu \to e \gamma$ \cite{ChangHouKeung};
$H^0 \to t {\bar c}$ \cite{Hou};
(tree level) process $\mu^+ \mu^- \to H^0 \to t {\bar c}$
\cite{AtwoodReinaSoni1};
$e^+e^- \to Z^{\ast} \to
H^0 A^0 \to b {\bar b} t {\bar c}$, $W^+W^- t {\bar c}$,
or $tt{\bar c} {\bar c}$ \cite{HouLin};
one-loop process $e^+e^- \to \gamma^{\ast}, Z^{\ast} \to
\to t {\bar c}$ \cite{AtwoodReinaSoni2},
\cite{AtwoodReinaSoni3}\footnote{
Ref.~\cite{AtwoodReinaSoni3}
contains, in addition, an analysis of constraints
from the electroweak $\rho$-parameter and
from experimental data of various low energy processes
such as $\triangle F = 2$ processes, $Z \to b {\bar b}$,
rare B decays, $e^+e^- \to b {\bar s}$.};
one-loop processes
$\gamma \gamma \to h^0, A^0 \to t {\bar c}$ \cite{HouLin};
$e^+ e^- \to Z^{\ast} \to (h^0, H^0)Z \to t {\bar c} Z$ \cite{HouLin};
gauge boson ($WW$ and $ZZ$)
fusion processes $e^+ e^- \to t {\bar c} \nu_e
{\bar \nu}_e$, $t {\bar c} e^+ e^-$ \cite{BarShalomEilamSoniWudka};
rare decays $t \to c W^+W^-$, $cZZ$ \cite{BarShalomEilamSoniWudka}.

Hall and Weinberg \cite{HallWeinberg}
pointed out that, while such 2HDM(III) models 
generically (naturally) suppress FCNC reaction rates to
acceptable levels (e.g., by possessing approximate
global flavor $U(1)$ symmetries leading to
AHR-type of ans\"atze), they may in general 
give too much CP violation in the neutral kaon mass
matrix. Therefore, they argued that these models
must also possess CP as a good approximate symmetry.

Low energy phenomenology of a special 2HDM(III) framework, 
in which the renormalized
FCNC Yukawa couplings (not involving the top quark) are suppressed
practically to zero and the additional FCNC contributions
come only from loops involving charged scalars, 
has also been investigated \cite{Cvetic}.

In the present work, we
construct RGE's for the Yukawa coupling parameters of quarks 
(and for quark masses) in the discussed 2HDM(III)
and thus obtain a means for investigations of the
high energy behavior of such theories.
The main motivation for the latter investigations
can be summarized in the following question:
For which (if any) phenomenologically acceptable ans\"atze
of the FCNC Yukawa coupling parameters at low energies
do we have a reasonable behavior of these parameters
at higher energies of evolution? Under the ``reasonable''
behavior we understand a rather tame evolution of
these parameters as the energy of probes increases
by several orders of magnitude. Stated otherwise,
this is the requirement that these 
parameters do not increase by order(s) of magnitude
in the region of the evolution energy which is not
very close (on the logarithmic scale) 
to the Landau pole\footnote{
Near the Landau pole of the top quark mass
the theory starts behaving generally
in a nonperturbative manner and the perturbative (one-loop)
RGE's start losing predictive power.}
of the top quark mass. Therefore,
the theory would not change qualitatively in the
FCNC Yukawa sector up until the energies of probes 
where the framework starts behaving nonperturbatively
due to the large ``mass'' Yukawa coupling of the top quark.
Such a reasonable behavior would then provide us with
additional\footnote{
-- additional to the {\it low energy} arguments of CSY and AHR.}
 arguments that the discussed 2HDM(III)
frameworks, at least for some of the specific
phenomenologically acceptable low energy choices of
Yukawa parameters, are not unnatural.

The Yukawa interactions 
%at energies $E$ of probes, 
in this 2HDM(III) framework in any $\mbox{SU(2)}_L$-basis 
have the most general form
\begin{eqnarray}
{\cal L}^{(E)}_{\mbox{\scriptsize Y(III)}} & = & 
- \sum_{i,j=1}^3 {\Big \lbrace}
{\tilde D}_{ij}^{(1)}( {\bar {\tilde q}}^{(i)}_L {\Phi}^{(1)} )
{\tilde d}^{(j)}_{R} +
{\tilde D}_{ij}^{(2)}( {\bar {\tilde q}}^{(i)}_L {\Phi}^{(2)} )
{\tilde d}^{(j)}_{R} +
    \nonumber\\
& & + {\tilde U}_{ij}^{(1)}( {\bar {\tilde q}}^{(i)}_L \tilde {\Phi}^{(1)} )
{\tilde u}^{(j)}_{R} +
{\tilde U}_{ij}^{(2)}( {\bar {\tilde q}}^{(i)}_L \tilde {\Phi}^{(2)} )
{\tilde u}^{(j)}_{R} + h.c.
{\Big \rbrace}
+ \lbrace  \ {\bar \ell} {\Phi} {\ell}\mbox{-terms} \ \rbrace \ .
\label{2HD30}
\end{eqnarray}
The tildes above the Yukawa coupling parameters
and above the quark fields means that these quantities are 
in an arbitrary $\mbox{SU(2)}_L$-basis 
(not in the mass basis).
The superscript $(E)$ for the Lagrangian density means that
the theory has a finite effective energy cutoff $E$,
and the reference to this evolution energy
$E$ was omitted at the fields and at the
Yukawa coupling parameters in order to have simpler notation
($E \sim 10^2$ GeV for renormalized quantities). 
The following notations are used:
\begin{equation}
{\Phi}^{(k)}  \equiv  { {\phi}^{(k)+} \choose {\phi}^{(k)0} } 
\equiv \frac{1}{\sqrt{2}} { {\phi}_1^{(k)} + i {\phi}_2^{(k)} \choose
         {\phi}_3^{(k)} + i {\phi}_4^{(k)} }  \ , \ 
{\tilde {\Phi}}^{(k)}  \equiv i {\tau}_2 {\Phi}^{(k)\ast} 
\equiv \frac{1}{\sqrt{2}} { {\phi}_3^{(k)} - i {\phi}_4^{(k)} \choose
         -{\phi}_1^{(k)} + i {\phi}_2^{(k)} }  \ , 
\label{2HDnot1}
\end{equation}
\begin{equation}
{\tilde q^{(i)}} = { {\tilde u^{(i)}} \choose {\tilde d^{(i)}} } \ :
\qquad
{\tilde q^{(1)}} = { {\tilde u} \choose {\tilde d} } \ , \
{\tilde q^{(2)}} = { {\tilde c} \choose {\tilde s} } \ , \
{\tilde q^{(3)}} = { {\tilde t} \choose {\tilde b} } \ ,
\label{2HDnot2}
\end{equation}
\begin{equation}
\langle {\Phi}^{(1)} \rangle_0 = 
\frac{e^{i\eta_1}}{\sqrt{2}} {0 \choose v_1} \ ,
\qquad
\langle {\Phi}^{(2)} \rangle_0 = \
\frac{e^{i\eta_2}}{\sqrt{2}} {0 \choose v_2} \ ,
\qquad v_1^2+v_2^2 = v^2 \ .
\label{2HDnot3}
\end{equation}
In (\ref{2HDnot3}), $v$ [$\equiv v(E)$] is the usual VEV 
needed for the electroweak symmetry breaking, {\it i.e.},
$v(E_{\mbox{\scriptsize ew}}) \approx
246$ GeV. The phase difference $\eta \equiv \eta_2 - \eta_1$ 
between the two VEV's 
in (\ref{2HDnot3}) may be nonzero; 
it represents CP violation originating from the
purely scalar 2HD sector (cf.~\cite{Gunionetal}).
The leptonic sector has been omitted in (\ref{2HD30}).

We note that the popular ``type I'' and ``type II'' models
are special cases (subsets)
of this framework, with part of FCNC Yukawa coupling parameters
being exactly zero
\begin{equation}
\mbox{2HDM(I):} \qquad
U^{(1)} = D^{(1)} = 0 \, \qquad
\mbox{2HDM(II):} \qquad
U^{(1)} = D^{(2)} = 0 \ .
\label{types}
\end{equation}
In the 2HDM(II), the family symmetry in 
${\cal {L}}_{\mbox{\scriptsize{Y}}}$
enforcing this complete FCNC Yukawa parameter suppression 
at {\it all} evolution energies
$E$ is of U(1)-type: 
$d^{(j)}_{R} \rightarrow \mbox{e}^{i \alpha} d^{(j)}_{R}$, 
${\Phi}^{(1)} \rightarrow \mbox{e}^{-i \alpha} {\Phi}^{(1)}$ (j=1,2,3),
the other fields remaining unchanged. This symmetry ensures that, 
in the course of renormalization, no loop-induced ($\ln \Lambda$
cutoff-dependent) Yukawa couplings other 
than those of the form of the 2HDM(II) can appear. 
In the 2HDM(I), the family symmetry is similar:
$d^{(j)}_R \rightarrow \mbox{e}^{i \alpha} d^{(j)}_R$,   
$ \ u^{(j)}_R \rightarrow \mbox{e}^{-i \alpha} u^{(j)}_R$, 
${\Phi}^{(1)} \rightarrow e^{-i \alpha} {\Phi}^{(1)}$. 
In contrast to type I and type II, in type III 2HDM's
there is no exact (family) symmetry enforcing the complete suppression
of the FCNC Yukawa coupling parameters. 
Stated otherwise, while FCNC Yukawa parameters
in this general framework, at least those not involving the
top quark, must be made quite small at low
energies of probes $E \sim E_{\mbox{\scriptsize ew}}$, 
they in general may increase
when the energy of probes $E$ increases. If they
increase by order(s) of magnitude in the energy
region well below the Landau pole, then
such models should be regarded as rather unnatural --
their behavior in the FCNC Yukawa sector 
is then drastically different at higher energies of probes
(not just very near the Landau pole) from the behavior at
low (electroweak) energies.

In Section 2, we discuss relations between various
notations for the scalar isodoublets and between various
bases of the quark fileds, and the ensuing changes in the
representation of the Yukawa coupling parameters -- in order
to highlight the suppression conditions imposed at
low evolution energies on the FCNC Yukawa coupling parameters
of the 2HDM(III).
In Section 3, we derive one-loop RGE's
of the Yukawa coupling parameters and of the scalar fields
(and hence of their VEV's)
in the described 2HDM(III) model, with the purpose
of investigating the behavior of the framework 
at evolving energies of probes $E$.
In Section 4, we show and discuss one typical numerical example
of the evolution of Yukawa coupling parameters in this
general framework, in particular the evolution of the
FCNC Yukawa parameters. The values of
these parameters at {\it low energies} 
($E \sim E_{\mbox{\scriptsize ew}}$) were chosen 
according to the CSY ansatz, and therefore they fulfill 
the FCNC suppression restrictions discussed in this Introduction
and in Section 2.
Section 5 is a summary of the results and conclusions.

\section{Conditions of FCNC suppression at low evolution energies}

The Lagrangian density (\ref{2HD30})\footnote{
Throughout this Section we omit, for simpler notation,
reference to the arbitrary evolution (cutoff) energy $E$ at the quark
fields, at the scalar fields and their VEV's and at the Yukawa
coupling parameters.}  
can be written in a form more convenient for consideration of
FCNC Yukawa coupling parameters by redefining the scalar
isodoublets in the following way:
\begin{eqnarray}
{\Phi}^{\prime(1)} & = & (\cos \beta) {\Phi}^{(1)} +
                         (\sin \beta) e^{-i {\eta}} {\Phi}^{(2)} \ ,
\nonumber\\
{\Phi}^{\prime(2)} & = & - (\sin \beta) {\Phi}^{(1)} +
                         (\cos \beta) e^{-i {\eta}} {\Phi}^{(2)} \ ,
\label{redefPhi}
\end{eqnarray}
where
\begin{equation}
\eta=\eta_2-\eta_1 
\label{xi}
\end{equation}
and
\begin{equation} 
\tan \beta = \frac{v_2}{v_1} \ \Rightarrow \
\cos \beta = \frac{v_1}{v} \ , \quad \sin \beta = \frac{v_2}{v} \ .
\label{ratioVEV}
\end{equation}
Therefore, the VEV's of the redefined scalar isodoublets are
\begin{equation}
e^{-i {\eta}_1} \langle {\Phi}^{\prime(1)} \rangle_0 = \frac{1}{\sqrt{2}}
{0 \choose v} \ , \qquad
\langle {\Phi}^{\prime(2)} \rangle_0 = \frac{1}{\sqrt{2}}
{0 \choose 0} \ .
\label{newVEVs}
\end{equation}
The isodoublet ${\Phi}^{\prime(1)}$ is therefore responsible
for the masses of the quarks, and ${\Phi}^{\prime(2)}$ with
its couplings to the quarks is
responsible for the FCNC couplings, as will be seen below.
The original Yukawa Lagrangian density (\ref{2HD30}) of 2HDM(III)
can then be rewritten 
in terms of these redefined scalar fields as
\begin{eqnarray}
{\cal L}^{(E)}_{\mbox{\scriptsize Y(III)}} & = & 
- \sum_{i,j=1}^3 {\Big \lbrace}
{\tilde G}^{(D)}_{ij}( {\bar {\tilde q}}^{(i)}_L {\Phi}^{\prime (1)} )
{\tilde d}^{(j)}_{R} +
{\tilde G}^{(U)}_{ij}( {\bar {\tilde q}}^{(i)}_L {\tilde \Phi}^{\prime(1)} )
{\tilde u}^{(j)}_{R} + \mbox{h.c. } {\Big \rbrace}
\nonumber\\
&&
- \sum_{i,j=1}^3 {\Big \lbrace}
{\tilde D}_{ij}( {\bar {\tilde q}}^{(i)}_L {\Phi}^{\prime (2)} )
{\tilde d}^{(j)}_{R} +
{\tilde U}_{ij}( {\bar {\tilde q}}^{(i)}_L {\tilde \Phi}^{\prime(2)} )
{\tilde u}^{(j)}_{R} + \mbox{h.c. } {\Big \rbrace} \ ,
\label{Lnew}
\end{eqnarray}
where the Yukawa matrices ${\tilde G}^{(U)}$ and ${\tilde G}^{(D)}$
are rescaled mass matrices, and ${\tilde U}$ and ${\tilde D}$
the corresponding ``complementary'' Yukawa matrices, in an
(arbitrary) $SU(2)_L$-basis
\begin{eqnarray}
{\tilde G}^{(U)} = \frac{\sqrt{2}}{v} {\tilde M}^{(U)} &=&
(\cos \beta) {\tilde U}^{(1)} + 
(\sin \beta) e^{-i {\eta}} {\tilde U}^{(2)} \ ,
\nonumber\\
{\tilde G}^{(D)} = \frac{\sqrt{2}}{v} {\tilde M}^{(D)} &=&
(\cos \beta) {\tilde D}^{(1)} + 
(\sin \beta) e^{+i {\eta}} {\tilde D}^{(2)} \ ;
\label{Gs}
\end{eqnarray}
\begin{eqnarray}
{\tilde U} & = & - (\sin \beta) {\tilde U}^{(1)} +
(\cos \beta) e^{-i{\eta}} {\tilde U}^{(2)} \ ,
\nonumber\\
{\tilde D} & = & - (\sin \beta) {\tilde D}^{(1)} +
(\cos \beta) e^{+i{\eta}} {\tilde D}^{(2)} \ .
\label{UDs}
\end{eqnarray}
By a biunitary transformation involving unitary matrices
$V_L^U$, $V_R^U$, $V_L^D$ and $V_R^D$, the Yukawa
parameters can be expressed in the mass basis of the quarks, 
where the (rescaled) mass matrices $G^{(U)}$ and $G^{(D)}$ are
diagonal and real
\begin{eqnarray}
G^{(U)} = \frac{\sqrt{2}}{v} M^{(U)}& = &
V_L^{U} {\tilde G}^{(U)} V_R^{U\dagger} \ , 
\quad M_{ij}^{(U)}= {\delta}_{ij} m_i^{(u)} \ ;
\nonumber\\
U & = & V_L^{U} {\tilde U} V_R^{U\dagger} \ ;
\label{GUUmass}
\\
G^{(D)} = \frac{\sqrt{2}}{v} M^{(D)}& =&
V_L^{D} {\tilde G}^{(D)} V_R^{D\dagger} \ ,
\quad M_{ij}^{(D)}= {\delta}_{ij} m_i^{(d)} \ ;
\nonumber\\
D & = & V_L^{D} {\tilde D} V_R^{D\dagger} \ ;
\label{GDDmass}
\end{eqnarray}
\begin{equation}
u_L = V_L^{U} {\tilde u}_L \ , \quad u_R = V_R^U {\tilde u}_R \ , \quad
d_L = V_L^{D} {\tilde d}_L \ , \quad d_R = V_R^D {\tilde d}_R \ .
\label{qmass}
\end{equation}
The lack of tildes above the Yukawa coupling parameters
and above the quark fields means that these quantities are in the
quark mass basis (at a given evolution energy $E$). The
Lagrangian density (\ref{Lnew}) can be written now in the
quark mass basis. 
The neutral current part of the Lagrangian density
in the quark mass basis is
\begin{eqnarray}
{\cal L}^{(E)}_{\mbox{\scriptsize{Y(III) neutral}}} &=&
- \frac{1}{\sqrt{2}} \sum_{i=1}^3  
{\Big \lbrace}
G^{(D)}_{ii} {\bar d}^{(i)}_L d^{(i)}_{R} 
( {\phi}^{\prime(1)}_3 + i {\phi}^{\prime(1)}_4 ) + 
\nonumber\\ 
&& + G^{(U)}_{ii} {\bar u}^{(i)}_L u^{(i)}_{R}
( {\phi}^{\prime(1)}_3 - i {\phi}^{\prime(1)}_4 )  
+ \mbox{h.c. } {\Big \rbrace}
\nonumber\\
&&
- \frac{1}{\sqrt{2}} \sum_{i,j=1}^3 {\Big \lbrace}
D_{ij} {\bar d}^{(i)}_L d^{(j)}_{R} 
( {\phi}^{\prime(2)}_3 + i {\phi}^{\prime(2)}_4 ) +  
\nonumber\\
&& + U_{ij} {\bar u}^{(i)}_L u^{(j)}_{R}
( {\phi}^{\prime(2)}_3 - i {\phi}^{\prime(2)}_4 )  
 + \mbox{h.c. } {\Big \rbrace} \ .
\label{Lmassn}
\end{eqnarray}
On the other hand, the charged current part of the Lagrangian density
in the quark mass basis is
\begin{eqnarray}
{\cal L}^{(E)}_{\mbox{\scriptsize{Y(III) charged}}} & =&  
- \frac{1}{\sqrt{2}} \sum_{i,j=1}^3  
{\Big \lbrace}
(V G^{(D)})_{ij} {\bar u}^{(i)}_L d^{(j)}_{R} 
( {\phi}^{\prime(1)}_1 + i {\phi}^{\prime(1)}_2 )  - 
\nonumber\\
&& - (V^{\dagger} G^{(U)})_{ij} {\bar d}^{(i)}_L u^{(j)}_{R}
( {\phi}^{\prime(1)}_1 - i {\phi}^{\prime(1)}_2 )  
+ \mbox{h.c. } {\Big \rbrace}
\nonumber\\
&&
- \frac{1}{\sqrt{2}} \sum_{i,j=1}^3 {\Big \lbrace}
(V D)_{ij} {\bar u}^{(i)}_L d^{(j)}_{R} 
( {\phi}^{\prime(2)}_1 + i {\phi}^{\prime(2)}_2 ) -
\nonumber\\ 
&&-(V^{\dagger} U)_{ij} {\bar d}^{(i)}_L u^{(j)}_{R}
( {\phi}^{\prime(2)}_1 - i {\phi}^{\prime(2)}_2 )  
 + \mbox{h.c. } {\Big \rbrace} \ .
\label{Lmassch}
\end{eqnarray}
Here, we denoted by $V$ the Cabibbo-Kobayashi-Maskawa (CKM) matrix
\begin{equation}
V \equiv V_{\mbox{\scriptsize CKM}} = V_L^U V_L^{D\dagger} \ .
\label{CKM}
\end{equation}
We see from (\ref{Lmassn}) that the $U$ and $D$ matrices,
as defined by (\ref{UDs}) and (\ref{GUUmass})-(\ref{GDDmass})
through the original Yukawa matrices ${\tilde U}^{(j)}$ and
${\tilde D}^{(j)}$ of the 2HDM(III) Lagrangian density
(\ref{2HD30}), allow the model to possess 
in general scalar-mediated
FCNC's. Namely, in the quark mass basis only
the (rescaled) quark mass matrices $G^{(U)}$ and $G^{(D)}$ 
of (\ref{GUUmass})-(\ref{GDDmass}) [cf.~also (\ref{Gs})]
are diagonal, but the matrices $U$ and $D$ in this
general framework are in general not diagonal
\begin{eqnarray}
{\cal L}^{(E)}_{\mbox{\scriptsize{Y(III) FCNC}}} & =&
- \frac{1}{2 \sqrt{2}} \sum_{
\vspace{-2.mm}
\begin{array}{c}
{\scriptstyle i,j = 1} \\[-2.mm]
{\scriptstyle i \not= j}
\end{array}}^3 {\Big [} \left( D+D^{\dagger} \right)_{ij}
\left( {\bar d}^{(i)} d^{(j)} \right) {\phi}_3^{\prime (2)} 
+ i \left( D - D^{\dagger} \right)_{ij} 
\left( {\bar d}^{(i)} d^{(j)} \right) {\phi}_4^{\prime (2)}
\nonumber\\
&& + \left( D+D^{\dagger} \right)_{ij}
\left( {\bar d}^{(i)} i {\gamma}_5 d^{(j)} \right) {\phi}_4^{\prime (2)} 
- i \left( D - D^{\dagger} \right)_{ij} 
\left( {\bar d}^{(i)} i {\gamma}_5 d^{(j)} \right) {\phi}_3^{\prime (2)}
{\Big ]}
\nonumber\\
&&
- \frac{1}{2 \sqrt{2}} \sum_{
\vspace{-2.mm}
\begin{array}{c}
{\scriptstyle i,j = 1} \\[-2.mm]
{\scriptstyle i \not= j}
\end{array}}^3 {\Big [} \left( U+U^{\dagger} \right)_{ij}
\left( {\bar u}^{(i)} u^{(j)} \right) {\phi}_3^{\prime (2)} 
- i \left( U - U^{\dagger} \right)_{ij} 
\left( {\bar u}^{(i)} u^{(j)} \right) {\phi}_4^{\prime (2)}
\nonumber\\
&& - \left( U+U^{\dagger} \right)_{ij}
\left( {\bar u}^{(i)} i {\gamma}_5 u^{(j)} \right) {\phi}_4^{\prime (2)} 
- i \left( U - U^{\dagger} \right)_{ij} 
\left( {\bar u}^{(i)} i {\gamma}_5 u^{(j)} \right) {\phi}_3^{\prime (2)}
{\Big ]} \ .
\label{FCNCs}
\end{eqnarray}

It should be noted that the original four Yukawa matrices
${\tilde U}^{(j)}$ and ${\tilde D}^{(j)}$ ($j=1,2$)
in an $SU(2)_L$-basis
are already somewhat constrained by the requirement that (at low
energy) the squares of the linear combinations\footnote 
{Strictly speaking, the following
``squares'': $M^{(U)} M^{(U)\dag}$ and $M^{(D)} M^{(D)\dag}$.} 
$M^{(U)}$ and $M^{(D)}$ are
diagonalized by unitary transformations
involving such unitary matrices $V_L^U$ and $V_L^D$,
respectively, which are related
to each other by $V_L^U V_L^{D\dag} = V$. Here, $V$ is the
CKM matrix which is, for any specific chosen
phase convention, more or less known at low energies.

In order to have at low evolution energies ($E \sim
E_{\mbox{\scriptsize ew}}$) a phenomenologically viable
suppression of the scalar-mediated FCNC's, the
authors Cheng, Sher and Yuan (CSY) 
\cite{ChengSher}, \cite{SherYuan} basically argued that
the elements of the $U$ and $D$ matrices (in the quark 
mass basis and at low evolution energies $E$) should have the form:
\begin{equation}
U_{ij}(E) = {\xi}^{(u)}_{ij} \frac{\sqrt{2}}{v}
\sqrt{m_i^{(u)} m_j^{(u)}} \ , \qquad
D_{ij}(E) = {\xi}^{(d)}_{ij} \frac{\sqrt{2}}{v}
\sqrt{m_i^{(d)} m_j^{(d)}} \ , 
\label{FCNCcon1}
\end{equation}
where
\begin{equation}
{\xi}^{(u)}_{ij}, {\xi}^{(d)}_{ij} \sim 1 \ \mbox{for } 
E \sim E_{\mbox{\scriptsize ew}} \ .
\label{FCNCcon2}
\end{equation}
This form is in general phenomenologically acceptable and is
motivated basically only by the actual
mass hierarchies of quarks and the requirement
that there is [at a given {\it low} energy of evolution
($\sim E_{\mbox{\scriptsize ew}}$)] no
fine-tuning in which large Yukawa terms
$U^{(i)}_{jk}$ (and: $D^{(i)}_{jk}$)
add together via (\ref{UDs})
to make small terms $U_{jk}$ ($D_{jk}$).\footnote{
To visualize this point, Eqs.~(\ref{Gs}) and (\ref{UDs})
should be inspected, but this time in the quark mass
basis ({\it i.e.}, no tildes over the matrices). Then
(\ref{Gs}) would suggest that $U^{(i)}_{jk}$
($D^{(i)}_{jk}$) is in general non-diagonal and of the order
of $\sqrt{m_j^{(u)} m_k^{(u)}}/v$
($\sqrt{m_j^{(d)} m_k^{(d)}}/v$). As a result,
(\ref{UDs}) would suggest that $U_{jk}$
($D_{jk}$) is also of that order of magnitude, unless
there is some peculiar fine-tuning on the 
right-hand side (RHS) of (\ref{UDs}).
For the complete suppression of FCNC Yukawa couplings
($U_{jk}=0=D_{jk}$ for $j \not= k$) we would
need fine-tuning on the RHS of (\ref{UDs}).}
Therefore, this (CSY) form is
considered to be reasonably natural. Similar (but not
identical) ans\"atze have been proposed by
the authors of \cite{AntaramianHallRasin} (AHR),
motivated by their requirement that the Yukawa interactions
have approximate $U(1)$ flavor symmetries.

{}From the CSY ansatz (\ref{FCNCcon1})-(\ref{FCNCcon2})
we see that the scalar-mediated FCNC vertices
involving the heavy top quark are the only ones that are not
strongly suppressed (at low evolution energies), since,
as mentioned in the Introduction, FCNC processes involving
the top quark vertices (not loops) are not constrained
by present experiments. Later in Section 4 we will
use low energy conditions (\ref{FCNCcon1})-(\ref{FCNCcon2}) for
a numerical example of RGE flow of FCNC Yukawa coupling
parameters.

\section{Renormalization Group Equations (RGE's) in the general 2HDM}

\subsection{RGE's for the scalar fields}

Here we present a derivation of the one-loop RGE's
for the scalar fields (\ref{2HDnot1}). The one-loop
RGE's for the Yukawa coupling matrices ${\tilde D}^{(k)}$,
${\tilde U}^{(k)}$ ($k=1,2$) will be presented in the
next Subsection. In both derivations we will follow the
finite-cutoff interpretation of RGE's as discussed, for example,
by Lepage \cite{Lepage}.

In order to calculate evolution of the scalar fields
${\phi}_i^{(k)}(E)$ with ``cutoff'' energy $E$,
we need to calculate first the truncated (one-loop) two-point Green
functions $-i \Sigma_{ij}^{(k,\ell)}(p^2; E^2)$ represented 
diagrammatically in Fig.~1.
\begin{figure}[htb]
\mbox{}
\vskip4.cm\relax\noindent\hskip3.2cm\relax
\includegraphics{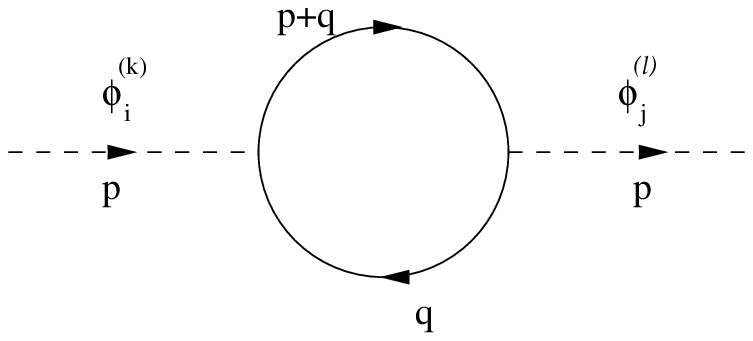} \vskip-0.5cm
\caption{\footnotesize The diagram leading to the
two-point Green function $-i \Sigma_{ij}^{(k,\ell)}(p^2; E^2)$.
Full lines represent quark propagators.}
\end{figure}
 More specifically, we calculate
their cutoff-dependent parts $\propto p^2 \ln E^2$ which
will be ultimately responsible for the effective\footnote{
For clearer notation, we denote in this Section the evolving
(UV cutoff) energy $E$ at the fields not as a superscript, but
rather as an argument.}
kinetic-energy-type
terms $\sim {\partial}_{\nu}{\phi}_i^{(k)}(E) 
{\partial}^{\nu} {\phi}_j^{(\ell)}(E)$. In the course of the
calculations, we ignore all the masses $m \sim E_{\mbox{\scriptsize ew}}$
of the relevant particles in the diagram. This would be consistent with
the picture of a framework with a finite but large ultraviolet
energy cutoff $E \gg E_{\mbox{\scriptsize ew}}$. For this reason,
we don't have to work in the mass basis of the relevant particles
-- these particles are regarded as effectively massless in the
approximation, the transformations between the original
bases of the relevant fields and their mass bases
are unitary, and therefore the (mass-independent parts of the)
calculated Green functions are the same in both bases.

Calculation of the mentioned two-point Green functions
$- i {\Sigma}_{i,j}^{(k,\ell)}(p^2; E^2)$,
whose external (scalar) legs ${\phi}_i^{(k)}$ and
${\phi}_j^{(\ell)}$ are truncated, is in the mentioned
framework rather straightforward. The relevant (massless) integrals
over internal quark-loop momenta $q$ can be carried out,
for example, in Euclidean metric [${\bar q} =
(-i q^0, -q^j)$, ${\bar p} = (- i p^0, -p^j)$], 
where the upper bound in the loop integral
is: ${\bar q}^2 \leq E^2$. After rotating back into
Minkowski metric (${\bar p}^2 \mapsto -p^2$), we end up
with the following results:\footnote{
Throughout this Section, we implicitly ignore the cutoff-independent
parts of the Green functions, since these parts are
irrelevant for the mass-independent (${\overline {MS}}$-type of)
RGE framework considered here.}
\begin{enumerate}
\item 
Green functions with the external legs ${\phi}_i^{(k)}$ and 
${\phi}_j^{(\ell)}$ having the same scalar indices ($i=j$):
\begin{eqnarray}
- i {\Sigma}_{j,j}^{(k,\ell)}(p^2;E^2) & = & i \frac{\kappa}{2}
p^2 \ln \left( \frac{E^2}{m^2} \right)
\nonumber\\
&& \times 
\mbox{Tr} \left[ {\tilde U}^{(k)} {\tilde U}^{(\ell) \dagger} +
{\tilde U}^{(\ell)} {\tilde U}^{(k) \dagger} +
{\tilde D}^{(k)} {\tilde D}^{(\ell) \dagger} +
{\tilde D}^{(\ell)} {\tilde D}^{(k) \dagger} 
\right] (E) \ ,
\label{Sigkk}
\end{eqnarray}
where $j = 1,2,3,4$ (no running over the repeated indices $j$); 
$k, \ell =1,2$; 
$m$ is an arbitrary but fixed mass of the
order of electroweak scale ($m \sim E_{\mbox{\scriptsize ew}}$); 
and ${\kappa}$ stands for
\begin{displaymath}
{\kappa} \equiv \frac{ N_{\mbox{\scriptsize c}} }{ 16 \pi^2} \ .
\end{displaymath}
\item
Green functions with the external legs ${\phi}_i^{(k)}$ and 
${\phi}_j^{(\ell)}$ having different scalar indices
($i \not= j$):
\begin{eqnarray}
\lefteqn{
- i {\Sigma}_{1,2}^{(1,2)}(p^2; E^2) =
- i {\Sigma}_{3,4}^{(1,2)}(p^2; E^2) }
\nonumber\\
& = &
- \frac{\kappa}{2} p^2 \ln \left( \frac{E^2}{m^2} \right)
\mbox{Tr} \left[ {\tilde U}^{(1)} {\tilde U}^{(2)\dagger} - 
{\tilde U}^{(2)} {\tilde U}^{(1)\dagger} - 
{\tilde D}^{(1)} {\tilde D}^{(2)\dagger} + 
{\tilde D}^{(2)} {\tilde D}^{(1)\dagger} \right] (E) \ .
\label{Sigkl1}
\end{eqnarray}
The above Green functions are antisymmetric under the
exchange of the Higgs generation indices $k \not= \ell$, 
and also under the exchange of the scalar indices
$i \not= j$
\begin{eqnarray}
&& - i {\Sigma}_{1,2}^{(1,2)} = + i {\Sigma}_{1,2}^{(2,1)}
= - i {\Sigma}_{2,1}^{(2,1)} = + i {\Sigma}_{2,1}^{(1,2)} 
\nonumber\\
& = &- i {\Sigma}_{3,4}^{(1,2)} = + i {\Sigma}_{3,4}^{(2,1)}
= - i {\Sigma}_{4,3}^{(2,1)} = + i {\Sigma}_{4,3}^{(1,2)} \ .
\label{Sigkl2}
\end{eqnarray}
\item
The other Green functions are zero
\begin{equation}
-i {\Sigma}_{i,j}^{(k,\ell)} = 0
\label{Sigother}
\end{equation}
for $i=1,2$ and $j=3,4$; for $i=3,4$ and $j=1,2$;
for $i \not= j$ and $k={\ell}$.
\end{enumerate}
All these Green functions can be induced at the tree level
by kinetic energy terms. For example, in theory with
the UV cutoff $E$, 
the kinetic energy term ${\partial}_{\nu} {\phi}_i^{(k)}(E)
{\partial}^{\nu} {\phi}_j^{(\ell)}(E)$ induces (at the tree level)
the two-point Green
function value $-i {\Sigma}_{i,j}^{(k,\ell)}(p^2; E^2) = i p^2$ 
if ${\phi}_i^{(k)}(E) \not= {\phi}_j^{(\ell)}(E)$, 
and the value $2 i p^2$ if 
${\phi}_i^{(k)}(E) \equiv {\phi}_j^{(\ell)}(E)$. 
Now, following the finite-cutoff interpretation of RGE's
as described, for example, by Lepage \cite{Lepage},
we compare the kinetic energy terms in the theory with
the UV cutoff $E$ and the equivalent theory with the slightly 
different cutoff $(E + dE)$. The two-point Green
functions in these two equivalent theories must be identical.
When imposing this requirement in the tree $+$ one-loop
approximation, this leads to the following relation:
\begin{eqnarray}
\lefteqn{
\frac{1}{2} \sum_{j=1}^4 
{\partial}_{\nu} \left[ \begin{array}{cc}
{\phi}_j^{(1)}, & {\phi}_j^{(2)} 
\end{array} \right](E)
\left[ \begin{array}{cc}
1 & 0 \\
0 & 1 
\end{array} \right]
{\partial}^{\nu} \left[ \begin{array}{c}
{\phi}_j^{(1)} \\
{\phi}_j^{(2)}
\end{array} \right](E)  =  }
\nonumber\\
&& 
\frac{1}{2} \sum_{j=1}^4 
{\partial}_{\nu} \left[ \begin{array}{cc}
{\phi}_j^{(1)}, & {\phi}_j^{(2)} 
\end{array} \right] (E + d E)
\left[ \begin{array}{cc}
1 & 0 \\
0 & 1 
\end{array} \right]
{\partial}^{\nu} \left[ \begin{array}{c}
{\phi}_j^{(1)} \\
{\phi}_j^{(2)}
\end{array} \right] (E + d E)  
\nonumber\\
&& + \frac{{\kappa}}{2} ( d \ln E^2 ) \sum_{j=1}^4 
{\partial}_{\nu} \left[ \begin{array}{cc}
{\phi}_j^{(1)}, & {\phi}_j^{(2)} 
\end{array} \right](E)
\left[ \begin{array}{cc}
A_{11}(E) & A_{12}(E) \\
A_{21}(E) & A_{22}(E)
\end{array} \right]
{\partial}^{\nu} \left[ \begin{array}{c}
{\phi}_j^{(1)} \\
{\phi}_j^{(2)}
\end{array} \right](E)
\nonumber\\
&& +  \frac{{\kappa}}{2} ( d \ln E^2 ) \sum_{(i,j)} (-1)^j 
{\partial}_{\nu} \left[ \begin{array}{cc}
{\phi}_i^{(1)}, & {\phi}_j^{(2)} 
\end{array} \right](E)
\left[ \begin{array}{cc}
0 & B_{12}(E) \\
B_{21}(E) & 0
\end{array} \right]
{\partial}^{\nu} \left[ \begin{array}{c}
{\phi}_i^{(1)} \\
{\phi}_j^{(2)}
\end{array} \right](E) \ ,
\label{kinLep}
\end{eqnarray}
where: the summation in the last sum runs over
$(i,j)=(1,2), (2,1), (3,4), (4,3)$; $d (\ln E^2) \equiv
\ln(E + d E)^2 - \ln E^2 =  2 d E/E$; and the real
matrix elements $A_{k \ell}$ and $B_{k \ell}$ are related to the
one-loop two-point Green function expressions
(\ref{Sigkk}) and (\ref{Sigkl1})-(\ref{Sigother}), respectively:
\begin{eqnarray}
A_{k \ell}(E) &=& \frac{1}{2} \mbox{Tr} \left[
{\tilde U}^{(k)} {\tilde U}^{(\ell)\dagger} +
{\tilde U}^{(\ell)} {\tilde U}^{(k)\dagger} +
{\tilde D}^{(k)} {\tilde D}^{(\ell)\dagger} +
{\tilde D}^{(\ell)} {\tilde D}^{(k)\dagger}  \right](E) \ ,
\label{As}
\\
B_{12}(E) & = &\frac{i}{2} \mbox{Tr} \left[
{\tilde U}^{(1)} {\tilde U}^{(2)\dagger} -
{\tilde U}^{(2)} {\tilde U}^{(1)\dagger} -
{\tilde D}^{(1)} {\tilde D}^{(2)\dagger} +
{\tilde D}^{(2)} {\tilde D}^{(1)\dagger}  \right](E)
= B_{21}(E) \ .
\label{Bs}
\end{eqnarray}
Equation (\ref{kinLep}) is described in the following way:
the sum on the left-hand side (LHS) 
and the first sum on the RHS represent 
the full kinetic energy terms of the
scalars in the formulation with the UV cutoff $E$ and $(E+ d E)$,
respectively. The one-loop contributions of Fig.~1
with the loop momentum $|{\bar q}|$ in the energy interval
$E \leq |{\bar q}| \leq {\Lambda}$ are already contained in the
kinetic energy terms of the LHS effectively at the tree level
(${\Lambda}$ is a large cutoff where the theory is presumed to
break down). On the other hand,
the kinetic energy terms of the $(E + d E)$
cutoff formulation [the first sum on the RHS of (\ref{kinLep})] 
effectively contain, at the tree level, the one-loop effects of Fig.~1
for the slightly smaller energy interval: $(E + d E) \leq
|{\bar q}| \leq {\Lambda}$. Therefore, the one-loop
two-point Green function contributions\footnote{
More precisely: the corresponding effective kinetic energy terms.}
$- i d {\Sigma}_{i,j}^{(k,\ell)}(p^2;E^2)$
of Fig.~1 from the loop-momentum interval 
$E \leq |{\bar q}| \leq (E + dE)$ had to be included
on the RHS of (\ref{kinLep}) -- these are the terms in the
last two sums there. This is illustrated in Fig.~2.
\begin{figure}[htb]
\mbox{}
\vskip3.cm\relax\noindent\hskip0.7cm\relax
\includegraphics{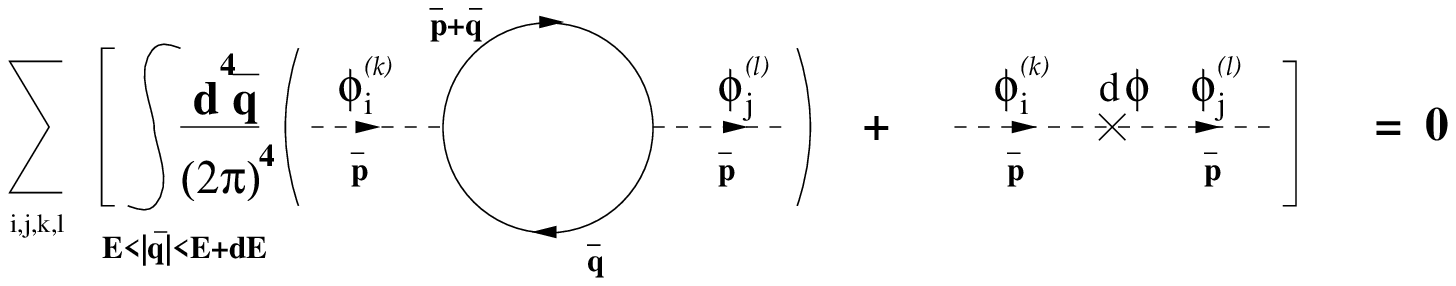} \vskip-0.9cm
\caption{\footnotesize Diagrammatic illustration of the
RGE relation (\ref{kinLep}) leading to the evolution of the
scalar fields. ${\phi}_i^{(k)}$ stands for ${\phi}_i^{(k)}(E)$,
and $d {\phi}$ stands for ${\phi}(E+dE)-{\phi}(E)$
($\phi$ is a generic notation for ${\phi}_j^{(\ell)}$'s). 
The cross represents the contribution of the change of the 
kinetic energy terms originating
from the changes $d {\phi}$ of scalar fields.}
\end{figure}

In order to find RGE's for the scalar fields ${\phi}_i^{(k)}(E)$,
we make the following ansatz for a solution of Eq.~(\ref{kinLep}):
\begin{equation}
{\vec \phi}_i(E+d E) 
\left\{ \equiv \left[ \begin{array}{c}
{\phi}_i^{(1)}\\
{\phi}_i^{(2)}
\end{array} \right] (E+d E) \right\}
= {\vec \phi}_i(E) +
{d \alpha}(i;E) {\vec \phi}_i(E) +
{d \beta}(i;E) {\vec \phi}_{i^{\prime}}(E) \ ,
\label{ansatz}
\end{equation}
where $d {\alpha}(i;E)$ and $d \beta(i;E)$ are infinitesimally
small $2 \times 2$ matrices
\begin{equation}
{d \alpha}(i;E) \equiv \left[
\begin{array}{cc}
{d \alpha}_{11} & {d \alpha}_{12} \\
{d \alpha}_{21} & {d \alpha}_{22} 
\end{array} \right] (i;E) \ ; \quad
{d \beta}(i;E) \equiv \left[
\begin{array}{cc}
0 & {d \beta}_{12} \\
{d \beta}_{21} & 0
\end{array} \right] (i;E) \ ,
\label{ansatznot}
\end{equation}
and the index $i^{\prime}$ in (\ref{ansatz}) is complementary to
index $i$:
\begin{displaymath}
i=1 \mapsto i^{\prime} = 2 \ , \quad
i=2 \mapsto i^{\prime} = 1 \ ; \qquad
i=3 \mapsto i^{\prime} = 4 \ , \quad
i=4 \mapsto i^{\prime} = 3 \ .
\end{displaymath}
Inserting ansatz (\ref{ansatz}) into the RGE relation (\ref{kinLep}),
we end up with the following set of relations:
\begin{equation}
{d \alpha}_{k \ell}(i;E) + {d \alpha}_{\ell k}(i;E) =
- {\kappa} \left( d \ln E^2 \right) A_{k \ell}(E) \ ,
\label{delalph}
\end{equation}
\begin{eqnarray}
\left[ {d \beta}_{12}(1;E) + {d \beta}_{21}(2;E) \right]
& =&
- \left[ {d \beta}_{21}(1;E) + {d \beta}_{12}(2;E) \right] =
- {\kappa} \left( d \ln E^2 \right) B_{12}(E) =
\nonumber\\
= \left[ {d \beta}_{12}(3;E) + {d \beta}_{21}(4;E) \right]&=&
- \left[ {d \beta}_{21}(3;E) + {d \beta}_{12}(4;E) \right] \ .
\label{delbet}
\end{eqnarray}
In principle, these relations alone do not define the
elements ${d \alpha}_{k \ell}(i;E)$ and
${d \beta}_{k \ell}(i;E)$. However, there is another
requirement that should be imposed on these transformation
coefficients: the resulting RGE evolution of the isodoublet fields
${\Phi}^{(1)}(E)$ and ${\Phi}^{(2)}(E)$ should be invariant
under the exchange of Higgs generation indices $1 \leftrightarrow 2$,
because these two Higgs doublets appear in the original
Lagrangian density (\ref{2HD30}) in a completely 
$1 \leftrightarrow 2$ symmetric manner. We will see in retrospect
that this discrete symmetry is respected once we impose the
conditions 
\begin{equation}
{d \alpha}_{k \ell}(i;E) = {d \alpha}_{\ell k}(i;E) \ ,
\label{symm1}
\end{equation}
\begin{eqnarray}
{d \beta}_{12}(1;E)& = & {d \beta}_{21}(2,E) \ , \qquad
{d \beta}_{12}(2;E) = {d \beta}_{21}(1,E) \ , 
\nonumber\\
{d \beta}_{12}(3;E)& = &{d \beta}_{21}(4,E) \ , \qquad
{d \beta}_{12}(4;E) = {d \beta}_{21}(3,E) \ .
\label{symm2}
\end{eqnarray}
Solutions (\ref{delalph})-(\ref{delbet}) of the RGE condition
(\ref{kinLep}), together with
the symmetry conditions (\ref{symm1})-(\ref{symm2}), lead to
specific expressions for the evolution
coefficients ${d \alpha}_{k \ell}(i;E)$
and ${d \beta}_{k \ell}(i;E)$. When inserting these coefficients
back into the scalar field evolution ansatz (\ref{ansatz}),
we end up with the following RGE's for the evolution of the scalar
fields:
\begin{eqnarray}
\lefteqn{
\frac{16 \pi^2}{N_{\mbox{\scriptsize c}}}
\frac{d}{d \ln E} {\phi}_j^{(1)}(E) =
 - \mbox{Tr} \left[ 
{\tilde U}^{(1)} {\tilde U}^{(1)\dagger} +
{\tilde D}^{(1)} {\tilde D}^{(1)\dagger} \right] {\phi}_j^{(1)} }
\nonumber\\
&&
- \frac{1}{2} \mbox{Tr} \left[ 
{\tilde U}^{(1)} {\tilde U}^{(2)\dagger} +
{\tilde U}^{(2)} {\tilde U}^{(1)\dagger} +
{\tilde D}^{(1)} {\tilde D}^{(2)\dagger} +
{\tilde D}^{(2)} {\tilde D}^{(1)\dagger}\right] {\phi}_j^{(2)} 
\nonumber\\
&&
+ i (-1)^j \frac{1}{2} \mbox{Tr} \left[ 
{\tilde U}^{(1)} {\tilde U}^{(2)\dagger} -
{\tilde U}^{(2)} {\tilde U}^{(1)\dagger} -
{\tilde D}^{(1)} {\tilde D}^{(2)\dagger} +
{\tilde D}^{(2)} {\tilde D}^{(1)\dagger}\right] {\phi}_{j^{\prime}}^{(2)} \ ,
\label{RGEphi1Y}
\end{eqnarray}
\begin{eqnarray}
\lefteqn{
\frac{16 \pi^2}{N_{\mbox{\scriptsize c}}}
\frac{d}{d \ln E} {\phi}_j^{(2)}(E) =
 - \mbox{Tr} \left[ 
{\tilde U}^{(2)} {\tilde U}^{(2)\dagger} +
{\tilde D}^{(2)} {\tilde D}^{(2)\dagger} \right] {\phi}_j^{(2)} }
\nonumber\\
&&
- \frac{1}{2} \mbox{Tr} \left[ 
{\tilde U}^{(1)} {\tilde U}^{(2)\dagger} +
{\tilde U}^{(2)} {\tilde U}^{(1)\dagger} +
{\tilde D}^{(1)} {\tilde D}^{(2)\dagger} +
{\tilde D}^{(2)} {\tilde D}^{(1)\dagger}\right] {\phi}_j^{(1)} 
\nonumber\\
&&
+ i (-1)^{j+1} \frac{1}{2} \mbox{Tr} \left[ 
{\tilde U}^{(1)} {\tilde U}^{(2)\dagger} -
{\tilde U}^{(2)} {\tilde U}^{(1)\dagger} -
{\tilde D}^{(1)} {\tilde D}^{(2)\dagger} +
{\tilde D}^{(2)} {\tilde D}^{(1)\dagger}\right] {\phi}_{j^{\prime}}^{(1)} \ ,
\label{RGEphi2Y}
\end{eqnarray}
where again $j^{\prime}$ is the scalar index complementary to index
$j$: $(j,j^{\prime}) = (1,2)$, $(2,1)$, $(3,4)$, $(4,3)$.
These RGE's lead to RGE's for scalar isodoublets
${\Phi}^{(k)}$:
\begin{equation}
\frac{16 \pi^2}{N_{\mbox{\scriptsize c}}}
\frac{d}{d \ln E} {\Phi}^{(k)}(E) =
 - \sum_{\ell = 1}^2 \mbox{Tr} \left[ 
{\tilde U}^{(k)} {\tilde U}^{(\ell)\dagger} +
{\tilde D}^{(\ell)} {\tilde D}^{(k)\dagger} \right] {\Phi}^{(\ell)}
\ .
\label{RGEPhiY}
\end{equation}
We see indeed that this set of one-loop RGE's is invariant under
the exchange $1 \leftrightarrow 2$, as required by the form of the
Yukawa Lagrangian density (\ref{2HD30}) of the 2HDM(III).

In addition to quark loops, there are also
loops of the electroweak gauge bosons
contributing to one-loop two-point Green functions of the
scalars. However, since these gauge bosons
couple to the Higgs isodoublets identically as in 
the minimal Standard Model (MSM), their contributions
to the RHS of RGE's (\ref{RGEphi1Y})-(\ref{RGEphi2Y})
and (\ref{RGEPhiY})
are the same as in the MSM.\footnote{
For these contributions
of EW gauge bosons in the MSM, see for example Arason {\em et al.}
\cite{Arasonetal}, App.~A. However, note that they use
for the $U(1)_Y$ gauge coupling $g_1$ a different, 
GUT-motivated, convention:
$(g_1^2)_{\mbox{\scriptsize Arason et al.}} = 
(5/3) (g_1^2)_{\mbox{\scriptsize here}}$. } 
Consequently, the full one-loop
RGE's for the evolution of the scalar isodoublets in 2HDM(III) are
\begin{eqnarray}
16 \pi^2 \frac{d}{d \ln E} {\Phi}^{(k)}(E) &=&
 - N_{\mbox{\scriptsize c}} \sum_{\ell = 1}^2 \mbox{Tr} \left[ 
{\tilde U}^{(k)} {\tilde U}^{(\ell)\dagger} +
{\tilde D}^{(\ell)} {\tilde D}^{(k)\dagger} \right] {\Phi}^{(\ell)}
\nonumber\\
&&+ \left[ \frac{3}{4} g_1^2(E) + \frac{9}{4} g_2^2(E) \right]
{\Phi}^{(k)}(E) \ .
\label{RGEPhi}
\end{eqnarray}
These RGE's are simultaneously also RGE's for the corresponding
VEV's (\ref{2HDnot3})
\begin{eqnarray}
16 \pi^2 \frac{d}{d \ln E} \left( e^{i \eta_k} v_k \right) & = &
 - N_{\mbox{\scriptsize c}} \sum_{\ell = 1}^2 \mbox{Tr} \left[ 
{\tilde U}^{(k)} {\tilde U}^{(\ell)\dagger} +
{\tilde D}^{(\ell)} {\tilde D}^{(k)\dagger} \right] 
\left( e^{i \eta_{\ell}} v_{\ell} \right)
\nonumber\\
&&+ \left[ \frac{3}{4} g_1^2(E) + \frac{9}{4} g_2^2(E) \right]
\left( e^{i \eta_k} v_k \right) \ .
\label{RGEVEVs}
\end{eqnarray}

In this paper we don't discuss the question of quadratic
cutoff terms ${\Lambda}^2$ which appear in the radiative
corrections to VEV's in any SM framework. In the MSM,
their consideration -- under the assumption of the top quark 
dominance of the radiative corrections in the scalar sector -- 
leads to severe upper bounds on the ultraviolet cutoff ${\Lambda}$
for a substantial subset of values of the
bare doublet mass and of the bare scalar self-interaction 
parameters $M^2({\Lambda})$ and $\lambda({\Lambda})$ --
cf.~Ref.~\cite{Fateloetal}.

\subsection{RGE's for the Yukawa matrices}

In order to derive one-loop RGE's for the Yukawa matrices ${\tilde U}^{(k)}$
and ${\tilde D}^ {(k)}$, we will need the results of the previous
Subsection concerning evolution of the scalar fields. In addition,
we will need evolution of the quark fields 
${\tilde u}^ {(j)}_{L,R}$ and ${\tilde d}^ {(j)}_{L,R}$. 
The latter can be derived in close analogy with the derivation
of the evolution of scalar fields of the previous Subsection.
Now, the diagrams (Green functions) of Figs.~1 and 2 are replaced by 
those of Figs.~3 and 4,
and the scalar field kinetic energy terms in (\ref{kinLep})
are replaced by those of the quark fields.
The one-loop two-point Green function of Fig.~3, 
\begin{figure}[htb]
\mbox{}
\vskip3.cm\relax\noindent\hskip2.2cm\relax
\includegraphics{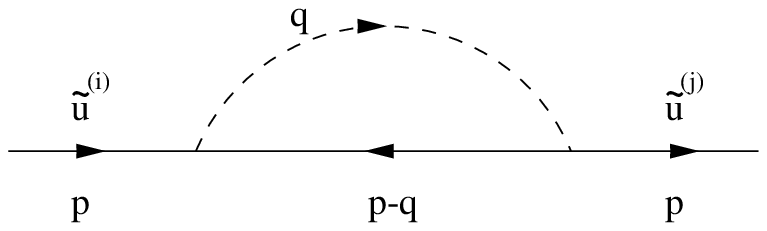} \vskip-0.1cm
\caption{\footnotesize The diagram leading to the
two-point Green function 
$-i \Sigma( p; E; {\tilde u}^{(i)}, {\tilde u}^{(j)} )$.
Dashed and full lines represent scalar and quark propagators, 
respectively.}
\end{figure}
with the
incoming ${\tilde u}^{(i)}$ and outgoing ${\tilde u}^{(j)}$
of momentum $p$,
in the framework with UV cutoff $E$, is
\begin{eqnarray}
- i {\Sigma} \left( 
p; E; {\tilde u}^ {(i)}, {\tilde u}^{(j)} \right) &=& 
\frac{i}{64 \pi^2} \ln \left( \frac{E^2}{m^2} \right) {p \llap /}
{\Bigg \{} \left(1 - {\gamma}_5 \right) \sum_{\ell=1}^2
\left[ {\tilde U}^{(\ell)} {\tilde U}^{(\ell)\dagger} +
{\tilde D}^{(\ell)} {\tilde D}^{(\ell)\dagger} \right]_{ji} 
\nonumber\\
&& + 2 \left(1+{\gamma}_5\right) \sum_{\ell=1}^2 
\left[ {\tilde U}^{(\ell)\dagger} {\tilde U}^{(\ell)} \right]_{ji}
{\Bigg \}} \ .
\label{Greenup}
\end{eqnarray}
The Green function with the incoming ${\tilde d}^{(i)}$
and outgoing ${\tilde d}^{(j)}$ of momentum $p$
is obtained from the
above expression by simply exchanging 
${\tilde U}^{(\ell)} \leftrightarrow {\tilde D}^{(\ell)}$ and
${\tilde U}^{(\ell)\dagger} \leftrightarrow {\tilde D}^{(\ell)\dagger}$.
We now make the ansatz for the running of the quark fields
\begin{eqnarray}
d{\tilde u}^{(k)}(E)_{L,R} 
{\Bigg \{} \equiv {\tilde u}^{(k)}(E+dE)_{L,R}-
{\tilde u}^{(k)}(E)_{L,R} {\Bigg \}}& = &
d f_u(E)_{k\ell}^{(L,R)}
{\tilde u}^{(\ell)}(E)_{L,R} \ , 
\label{ansup}
\\
d{\tilde d}^{(k)}(E)_{L,R} 
{\Bigg \{} \equiv {\tilde d}^{(k)}(E+dE)_{L,R}-
{\tilde d}^{(k)}(E)_{L,R} {\Bigg \}}& = &
d f_d(E)_{k\ell}^{(L,R)}
{\tilde d}^{(\ell)}(E)_{L,R} \ , 
\label{ansdown}
\end{eqnarray}
where subscripts $L$, $R$ denote handedness
of the quark fields: ${\tilde q}_L \equiv
(1 - {\gamma}_5) {\tilde q} / 2$, 
${\tilde q}_R \equiv (1 + {\gamma}_5 ) {\tilde q} / 2$
(${\tilde q} = {\tilde u}^{(k)}, {\tilde d}^{(k)}$).
In complete analogy with the previous Subsection,
we obtain from these ans\"atze and from the RGE relation
illustrated in Fig.~4\footnote{
The RGE relation represented by Fig.~4 is analogous
to relation (\ref{kinLep}) represented diagrammatically
in Fig.~2, but this time the kinetic energy terms are
those of the quark fields.} 
\begin{figure}[htb]
\mbox{}
\vskip3.cm\relax\noindent\hskip0.7cm\relax
\includegraphics{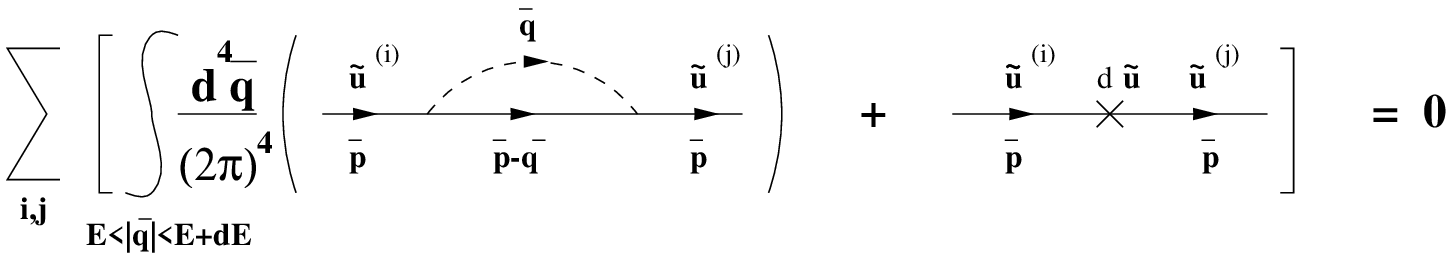} \vskip-1.1cm
\caption{\footnotesize Diagrammatic illustration of the
RGE relation leading to the evolution of quark fields.
Physically, this relation means that the two-point Green
functions with truncated external (quark) legs,
at one-loop level, are the same in the theory with
$E$ cutoff and in the theory with $E+dE$ cutoff.
Conventions are the same as in previous figures.}
\end{figure}
the following relations
for the quark field evolution matrices $d f_u$:
\begin{eqnarray}
d f_u(E)_{ij}^{(L)\ast} + d f_u(E)_{ji}^{(L)} & = &
- \frac{1}{32 \pi^2} (d \ln E^2) \sum_{k=1}^2 \left[
{\tilde U}^{(k)} {\tilde U}^{(k)\dagger} + 
{\tilde D}^{(k)} {\tilde D}^{(k)\dagger} \right]_{ji}(E) \ ,
\label{fuL}
\\
d f_u(E)_{ij}^{(R)\ast} + d f_u(E)_{ji}^{(R)} & = &
- \frac{2}{32 \pi^2} (d \ln E^2) \sum_{k=1}^2 \left[
{\tilde U}^{(k)\dagger} {\tilde U}^{(k)} \right]_{ji}(E) \ .
\label{fuR}
\end{eqnarray}
The relations for the $d f_d$ evolution matrices of the down-type
sector are obtained from the above by simple exchanges
${\tilde U}^{(\ell)} \leftrightarrow {\tilde D}^{(\ell)}$ and
${\tilde U}^{(\ell)\dagger} \leftrightarrow {\tilde D}^{(\ell)\dagger}$.
A solution to all these relations for the 
quark field evolution matrices is
\begin{eqnarray}
d f_u(E)_{ij}^{(L)} & = & 
- \frac{1}{64 \pi^2} (d \ln E^2) \sum_{k=1}^2 \left[
{\tilde U}^{(k)} {\tilde U}^{(k)\dagger} + 
{\tilde D}^{(k)} {\tilde D}^{(k)\dagger} \right]_{ij}(E) =
d f_d(E)_{ij}^{(L)} \ ,
\label{fLres}
\\
d f_u(E)_{ij}^{(R)} & = & 
- \frac{2}{64 \pi^2} (d \ln E^2) \sum_{k=1}^2 \left[
{\tilde U}^{(k)\dagger} {\tilde U}^{(k)} \right]_{ij}(E) \ ,
\label{fuRres}
\\
d f_d(E)_{ij}^{(R)} & = & 
- \frac{2}{64 \pi^2} (d \ln E^2) \sum_{k=1}^2 \left[
{\tilde D}^{(k)\dagger} {\tilde D}^{(k)} \right]_{ij}(E) \ .
\label{fdRres}
\end{eqnarray}

Another Green function needed for the derivation of the
one-loop RGE's of Yukawa matrices is the one represented
by the diagram of Fig.~5.
\begin{figure}[htb]
\mbox{}
\vskip4.5cm\relax\noindent\hskip-.8cm\relax
\includegraphics{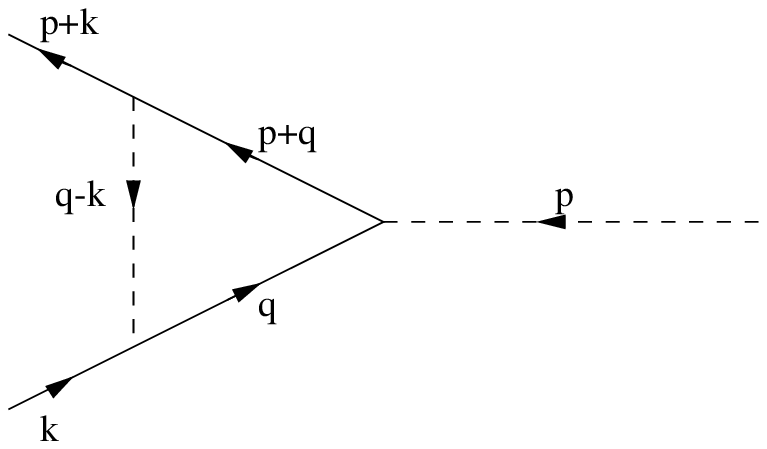} \vskip-0.1cm
\caption{\footnotesize One-particle-irreducible (1PI) diagram
contributing to the evolution of the Yukawa coupling
parameters. Conventions are the same is in previous
figures.}
\end{figure} 
When the external legs there are ${\tilde u}^{(i)}$ 
(incoming, with momentum $k$),
${\tilde u}^{(j)}$ 
(outgoing, with momentum $p+k$), and ${\phi}_3^{(\ell)}$
[or ${\phi}_4^{(\ell)}$], it turns out that only the
diagram with the {\em charged} scalar exchange
contributes, and the resulting truncated three-point Green function,
in the framework with the UV cutoff $E$, is
\begin{eqnarray}
\lefteqn{
G^{(3)}\left( k,p;E; {\tilde u}^{(i)},
{\tilde u}^{(j)}; {\phi}_3^{(\ell)}
\right)  =  - \frac{i}{32 \pi^2 \sqrt{2}}
\ln \left( \frac{E^2}{m^2} \right)  }
\nonumber\\
&& \times \sum_{r=1}^{2}
{\Bigg \{} (1+ {\gamma}_5) \left[ 
{\tilde D}^{(r)} {\tilde D}^{(\ell)\dagger} {\tilde U}^{(r)} 
\right]_{ji} +
(1 - {\gamma}_3) \left[
{\tilde U}^{(r)\dagger} {\tilde D}^{(\ell)} {\tilde D}^{(r)\dagger} 
\right]_{ji} {\Bigg \}} \ .
\label{3ptupGreen}
\end{eqnarray}
The corresponding Green function with the down-type quark
external legs is obtained from the above expression
by the simple exchanges
${\tilde U}^{(s)} \leftrightarrow {\tilde D}^{(s)}$ and
${\tilde U}^{(s)\dagger} \leftrightarrow {\tilde D}^{(s)\dagger}$.

Now, the one-loop RGE's for the Yukawa matrices are
obtained in analogy with the reasoning leading, in
the case of two-point scalar Green functions, to the
RGE relation (\ref{kinLep}) in the previous Subsection
[cf.~Fig.~2].
It is straightforward to check that the contribution
of the quark loops in the scalar external leg cancel
the contributions coming from the renormalizations
of the scalar fields in the kinetic energy terms
of the scalars -- this is illustrated in Fig.~6.
\begin{figure}[htb]
\mbox{}
\vskip3.cm\relax\noindent\hskip0.5cm\relax
\includegraphics{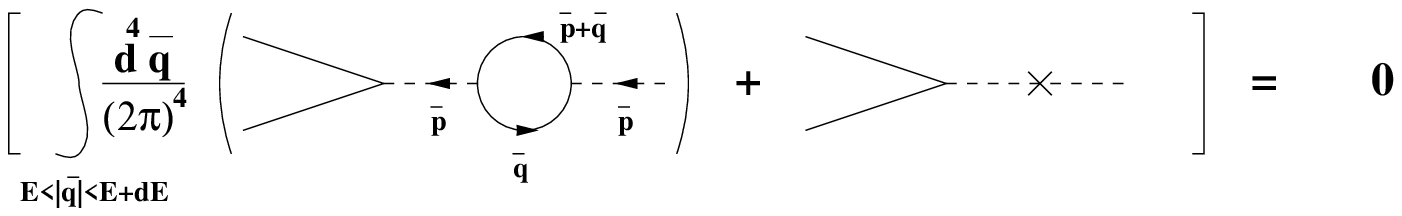} \vskip-0.9cm
\caption{\footnotesize Cancelation of contributions
from the quark loop
(one-particle-reducible -- 1PR) with those of the 
scalar field renormalizations in the kinetic
energy term of the scalars, for the energy cutoff
interval $(E,E+dE)$.}
\end{figure}
Furthermore, it can be checked
that the contributions of the scalar
exchanges on the external quark legs cancel the
contributions coming from the renormalizations
of the quark fields in the kinetic energy terms of
the quarks -- this is illustrated in Fig.~7.
\begin{figure}[htb]
\mbox{}
\vskip3.cm\relax\noindent\hskip0.3cm\relax
\includegraphics{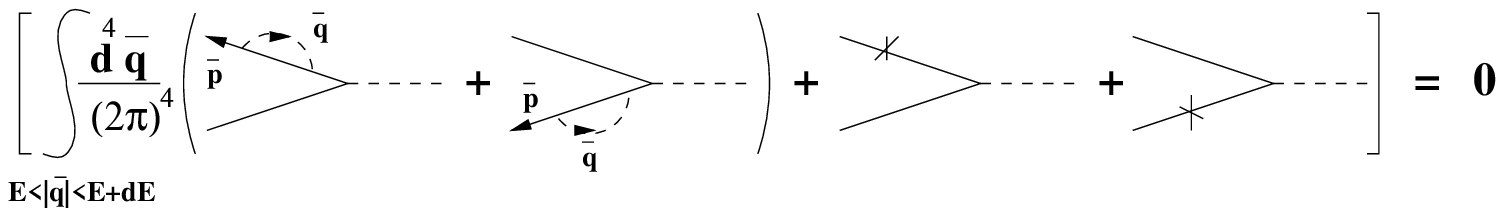} \vskip-1.0cm
\caption{\footnotesize Cancelation of contributions
from the scalar exchange on the quark legs
(1PR) with those of the 
quark field renormalizations in the kinetic
energy term of the quarks, for the energy cutoff
interval $(E,E+dE)$.}
\end{figure}
All in all, the 1PR 
one-loop contributions are canceled by the
contributions of field renormalizations in the kinetic energy
terms. Therefore, the only one-loop terms contributing
to the evolution of the ${\tilde U}^{(k)}$ Yukawa
matrices are those depicted in Fig.~8. 
\begin{figure}[htb]
\mbox{}
\vskip6.5cm\relax\noindent\hskip0.1cm\relax
\includegraphics{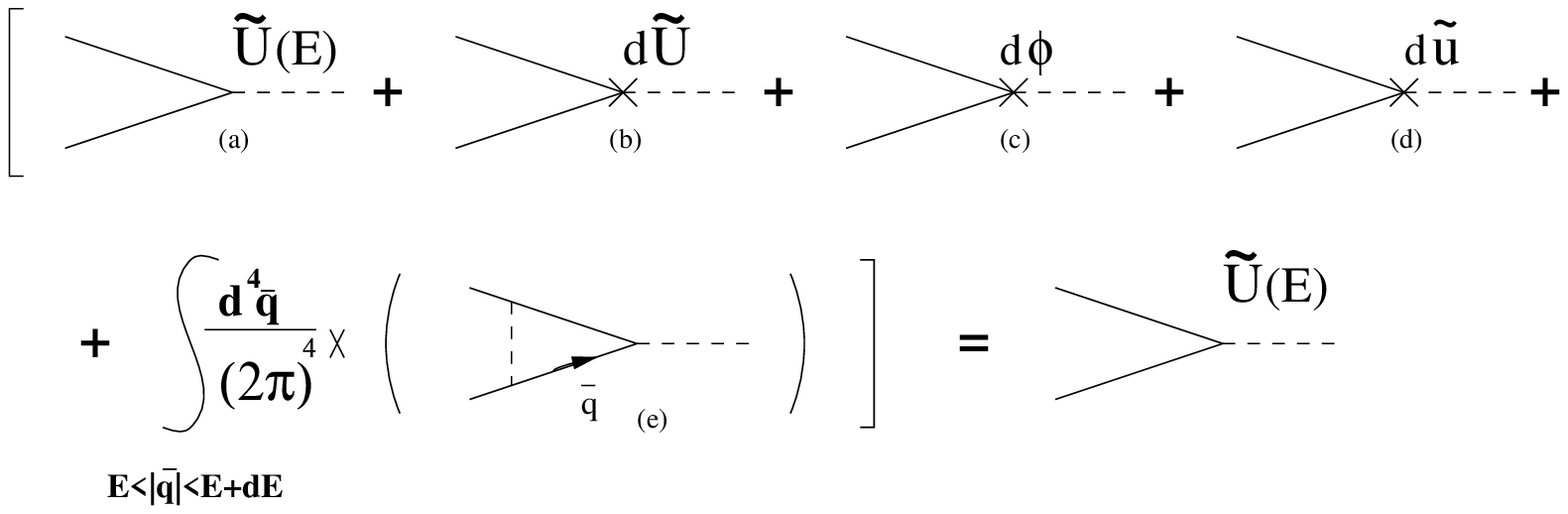} \vskip-0.9cm
\caption{\footnotesize Diagrammatic representation of the
RGE for the up-type Yukawa matrix ${\tilde U}$. Only
the 1PI scalar exchange [(e)]
and the effects of the renormalizations
of the Yukawa matrix, of the scalar fields and the quark fields
in the Yukawa couplings [(b), (c), (d), respectively] 
contribute when the cutoff is changed
from $E$ (RHS) to $E+dE$ (LHS). 
Note that $d{\tilde U}$ 
stands for ${\tilde U}(E+dE)- {\tilde U}(E)$,
{\it etc.} The contributions of the gauge boson exchanges were
not considered in the Figure.}
\end{figure}
The three
diagrams with crosses there correspond to contributions
of the following changes {\em in the Yukawa coupling} terms:
\begin{itemize}
\item
Yukawa matrix change (renormalization)
$d {\tilde U}^{(k)}$ [$\equiv {\tilde U}^{(k)}(E+dE) -
{\tilde U}^{(k)}(E)$] -- Fig.~8(b);
\item 
the scalar field renormalization $d {\tilde {\phi}}^{(k)}_s$
[$\equiv {\tilde {\phi}}^{(k)}_s(E+dE) -
{\tilde {\phi}}^{(k)}_s(E)$] -- Fig.~8(c);
\item
the quark field renormalization $d {\tilde u}^{(i)}$
and $d {\tilde u}^{(j)}$ -- Fig.~8(d).
\end{itemize}
Figure 8 is a diagrammatical representation of the physical
requirement that the three-point (quark-antiquark-scalar)
Green function, at one-loop level, be in the theory
with the cutoff $E+dE$ [left-hand side of Fig.~8: $(a)+\ldots +(e)$] 
the same as it is in the theory with the slightly lower cutoff $E$
(right-hand side).

Using the results of this and the previous Subsection,
we can then write down the one-loop RGE for
${\tilde U}^{(k)}$ corresponding to Fig.~8,
at the right-handed component [$ \propto (1+{\gamma}_5)$]
of the three-point Green function
\begin{eqnarray}
\lefteqn{
{\tilde U}^{(k)}_{ji} + d{\tilde U}^{(k)}_{ji}
+ \frac{1}{32 \pi^2} \left( d \ln E^2 \right)
{\Bigg \{} 
- N_{\mbox{\scriptsize c}} \sum_{\ell=1}^2 \mbox{Tr}
\left[ {\tilde U}^{(k)} {\tilde U}^{(\ell)\dagger} +
 {\tilde D}^{(\ell)} {\tilde D}^{(k)\dagger} \right]
{\tilde U}^{(\ell)} }
\nonumber\\
&& - \frac{1}{2} \sum_{\ell=1}^2 \left[
\left( {\tilde U}^{(\ell)} {\tilde U}^{(\ell)\dagger} +
{\tilde D}^{(\ell)} {\tilde D}^{(\ell)\dagger} \right)
{\tilde U}^{(k)} + 2 {\tilde U}^{(k)} {\tilde U}^{(\ell) \dagger}
{\tilde U}^{(\ell)} \right]
\nonumber\\
&&+ 2 \sum_{\ell=1}^2  
\left[ {\tilde D}^{(\ell)} {\tilde D}^{(k)\dagger} {\tilde U}^{(\ell)} 
\right]
{\Bigg \}}_{ji}
 =  {\tilde U}^{(k)}_{ji} \ .
\label{RGEU1}
\end{eqnarray}
The first sum on the LHS ($\propto N_{\mbox{\scriptsize c}}$)
corresponds to Fig.~8(c) [cf.~Eq.~(\ref{RGEPhiY})], 
the second sum to Fig.~8(d) [cf.~Eqs.~(\ref{fLres}), 
(\ref{fuRres})], 
and the third sum to Fig.~8(e) [cf.~Eq.~(\ref{3ptupGreen})].
The left-handed part of the Green function yields just the
Hermitean conjugate of the above matrix relation. The
analogous consideration of the three-point Green functions
with the down-type external quark legs ${\tilde d}^{(i)}$
and ${\tilde d}^{(j)}$ gives relations which can be
obtained from the above relation again by the simple
exchanges
${\tilde U}^{(s)} \leftrightarrow {\tilde D}^{(s)}$ and
${\tilde U}^{(s)\dagger} \leftrightarrow {\tilde D}^{(s)\dagger}$.
These relations can be rewritten in a more conventional
form
\begin{eqnarray}
16 \pi^2 \frac{d}{d \ln E} {\tilde U}^{(k)}(E) & = &
{\Bigg \{} 
N_{\mbox{\scriptsize c}} \sum_{\ell=1}^2
\mbox{Tr} \left[ {\tilde U}^{(k)} {\tilde U}^{(\ell)\dagger} +
 {\tilde D}^{(\ell)} {\tilde D}^{(k)\dagger} \right]
{\tilde U}^{(\ell)}
\nonumber\\
&&+ \frac{1}{2} \sum_{\ell=1}^2 \left[
{\tilde U}^{(\ell)} {\tilde U}^{(\ell)\dagger} +
{\tilde D}^{(\ell)} {\tilde D}^{(\ell)\dagger} \right] {\tilde U}^{(k)}
+ {\tilde U}^{(k)} \sum_{\ell=1}^2
{\tilde U}^{(\ell)\dagger} {\tilde U}^{(\ell)} 
\nonumber\\
&&-2 \sum_{\ell=1}^2 \left[
{\tilde D}^{(\ell)} {\tilde D}^{(k)\dagger}{\tilde U}^{(\ell)}
\right] {\Bigg \}} \ ,
\label{RGEU2}
\end{eqnarray}
and an analogous RGE for ${\tilde D}^{(k)}$.
These RGE's still don't contain one-loop effects of
exchanges of gauge bosons. However, since the couplings
of quarks and the Higgs doublets to the gauge bosons are
identical to those in the usual MSM, 2HDM(I) and 2HDM(II), 
their contributions
on the RHS of the above RGE's are identical to those in 
these theories. Therefore, the final form of the one-loop
RGE's for the Yukawa matrices in the general 2HDM(III)
now reads
\begin{eqnarray}
16 \pi^2 \frac{d}{d \ln E} {\tilde U}^{(k)}(E) & = &
{\Bigg \{} 
N_{\mbox{\scriptsize c}} \sum_{\ell=1}^2
\mbox{Tr} \left[ {\tilde U}^{(k)} {\tilde U}^{(\ell)\dagger} +
 {\tilde D}^{(\ell)} {\tilde D}^{(k)\dagger} \right]
{\tilde U}^{(\ell)}
\nonumber\\
&&+\frac{1}{2} \sum_{\ell=1}^2 \left[
{\tilde U}^{(\ell)} {\tilde U}^{(\ell)\dagger} +
{\tilde D}^{(\ell)} {\tilde D}^{(\ell)\dagger} \right] {\tilde U}^{(k)}
+ {\tilde U}^{(k)} \sum_{\ell=1}^2
{\tilde U}^{(\ell)\dagger} {\tilde U}^{(\ell)}
\nonumber\\ 
&&-2 \sum_{\ell=1}^2 \left[
{\tilde D}^{(\ell)} {\tilde D}^{(k)\dagger}{\tilde U}^{(\ell)}
\right]  - A_U {\tilde U}^{(k)}
{\Bigg \}} \ ,
\label{RGEUk}
\end{eqnarray}
\begin{eqnarray}
16 \pi^2 \frac{d}{d \ln E} {\tilde D}^{(k)}(E) & = &
{\Bigg \{} 
N_{\mbox{\scriptsize c}} \sum_{\ell=1}^2
\mbox{Tr} \left[ {\tilde D}^{(k)} {\tilde D}^{(\ell)\dagger} +
 {\tilde U}^{(\ell)} {\tilde U}^{(k)\dagger} \right]
{\tilde D}^{(\ell)}
\nonumber\\
&&+\frac{1}{2} \sum_{\ell=1}^2 \left[
{\tilde U}^{(\ell)} {\tilde U}^{(\ell)\dagger} +
{\tilde D}^{(\ell)} {\tilde D}^{(\ell)\dagger} \right] {\tilde D}^{(k)}
+ {\tilde D}^{(k)} \sum_{\ell=1}^2
{\tilde D}^{(\ell)\dagger} {\tilde D}^{(\ell)} 
\nonumber\\
&&-2 \sum_{\ell=1}^2 \left[
{\tilde U}^{(\ell)} {\tilde U}^{(k)\dagger}{\tilde D}^{(\ell)}
\right]  - A_D {\tilde D}^{(k)}
{\Bigg \}} \ ,
\label{RGEDk}
\end{eqnarray}
where the functions $A_U$ and $A_D$, characterizing the
contributions of the gauge boson exchanges, are the
same as in the MSM, 2HDM(I) and 2HDM(II)
\begin{eqnarray}
A_U & = & 
3 \frac{(N_{\mbox{\scriptsize c}}^2 - 1)}
{ N_{\mbox{\scriptsize c}} } g_3^2 + \frac{9}{4} g_2^2 +
\frac{17}{12} g_1^2 \ ,
\nonumber\\
A_D & = & A_U - g_1^2 \ ,
\label{AUAD}
\end{eqnarray}
and the gauge coupling parameters $g_j$ satisfy the one-loop
RGE's
\begin{equation}
16 \pi^2 \frac{d}{d \ln E} g_j= - C_j g_j^3 \ ,
\label{RGEgj}
\end{equation}
with the coefficients $C_j$ being those for the 2HDM's ($N_H = 2$)
\begin{equation}
C_3 = \frac{1}{3}(11 N_{\mbox{\scriptsize c}} - 2 n_q) \ ,
\quad C_2 = 7 - \frac{2}{3} n_q \ , \quad
C_1 = - \frac{1}{3} - \frac{10}{9} n_q \ .
\label{Cjs}
\end{equation}
Here, $n_q$ is the number of effective quark flavors --
e.g., for $E > m_t$ we have $n_q \approx 6$; for $m_b < E < m_t$
we have $n_q \approx 5$, etc.

The obtained relevant set of RGE's for the VEV's $e^{i \eta_k} v_k$
(\ref{RGEVEVs}) and the Yukawa matrices 
${\tilde U}^{(k)}$ and ${\tilde D}^{(k)}$ 
(\ref{RGEUk})-(\ref{RGEDk}) determines
in principle also the running of the quark masses.
Instead, we can rewrite all these RGE's in a representation involving
the VEV parameters $v \equiv \sqrt{v_1^2 + v_2^2}$, 
$\tan \beta \equiv v_2/v_1$ and 
$\eta \equiv \eta_2-\eta_1$
[cf.~(\ref{2HDnot3})], and matrices ${\tilde G}^{(U)}$,
${\tilde G}^ {(D)}$, ${\tilde U}$ and ${\tilde D}$
[cf.~(\ref{Gs}), (\ref{UDs})] -- this representation
is more convenient for discerning
the running of the quark masses and of the FCNC couplings.
Applying lengthy, but straightforward, algebra to the
hitherto obtained RGE's results in the RGE's of the
latter set of parameters
\begin{eqnarray}
16 \pi^2 \frac{d}{d \ln E} \left( v^2 \right) & = &
- 2 N_{\mbox{\scriptsize c}} \mbox{Tr} \left[
{\tilde G}^{(U)} {\tilde G}^ {(U)\dagger} +
{\tilde G}^{(D)} {\tilde G}^ {(D)\dagger} \right] v^2
+ \left[ \frac{3}{2} g_1^2 + \frac{9}{2} g_2^2 \right] v^2 \ ,
\label{RGEv2}
\end{eqnarray}
\begin{eqnarray}
16 \pi^2 \frac{d}{d \ln E} (\tan \beta) & = &
- \frac{N_{\mbox{\scriptsize c}}}{2 \cos^2 \beta} 
{\Bigg \{} \mbox{Tr} \left[{\tilde U} {\tilde G}^{(U)\dagger}
+ {\tilde G}^{(D)} {\tilde D}^ {\dagger} \right]
+ \mbox{Tr} \left[ {\tilde G}^{(U)} {\tilde U}^{\dagger}
+ {\tilde D} {\tilde G}^ {(D)\dagger} \right] {\Bigg \}} ,
\label{RGEtbeta}
\end{eqnarray}
\begin{eqnarray}
16 \pi^2 \frac{d}{d \ln E} (\eta) & = &
\frac{N_{\mbox{\scriptsize c}}}{i \sin (2 \beta)}
{\Bigg \{} \mbox{Tr} \left[ {\tilde G}^{(U)} {\tilde U}^{\dagger}
- {\tilde U} {\tilde G}^ {(U)\dagger} \right] %{\Bigg \}} 
-\mbox{Tr} \left[{\tilde G}^{(D)} {\tilde D}^{\dagger}
- {\tilde D} {\tilde G}^{(D)\dagger} \right] {\Bigg \}} \ ,
\label{RGExi}
\end{eqnarray}
\begin{eqnarray}
\lefteqn{
16 \pi^2 \frac{d}{d \ln E} ({\tilde U}) =
N_{\mbox{\scriptsize c}} {\Bigg \{}
2 \mbox{Tr} \left[ {\tilde U} {\tilde G}^{(U)\dagger}
+ {\tilde G}^{(D)} {\tilde D}^{\dagger} \right] {\tilde G}^ {(U)}
+ \mbox{Tr} \left[ {\tilde U} {\tilde U}^ {\dagger} +
{\tilde D} {\tilde D}^ {\dagger} \right] {\tilde U}
 }
\nonumber\\
&&
- \frac{1}{2} (\cot \beta) 
\mbox{Tr} \left[ {\tilde G}^{(U)} {\tilde U}^{\dagger}
- {\tilde U} {\tilde G}^{(U)\dagger} \right] {\tilde U}
+ \frac{1}{2} (\cot \beta) 
\mbox{Tr} \left[ {\tilde G}^{(D)} {\tilde D}^{\dagger}
- {\tilde D} {\tilde G}^{(D)\dagger} \right] {\tilde U} {\Bigg \}}
\nonumber\\
&& + {\Bigg \{} \frac{1}{2} \left[ {\tilde U} {\tilde U}^{\dagger}
+ {\tilde D} {\tilde D}^{\dagger} + {\tilde G}^{(U)} {\tilde G}^ {(U)\dagger}
+ {\tilde G}^{(D)} {\tilde G}^ {(D)\dagger} \right] {\tilde U} +
{\tilde U} \left[ {\tilde U}^{\dagger} {\tilde U} +
{\tilde G}^{(U)\dagger} {\tilde G}^{(U)} \right] 
\nonumber\\
&&
 - 2 {\tilde D} {\tilde D}^{\dagger} {\tilde U}
- 2 {\tilde G}^{(D)} {\tilde D}^ {\dagger} {\tilde G}^ {(U)} 
- A_U {\tilde U} {\Bigg \}} \ ,
\label{RGEU}
\end{eqnarray}
\begin{eqnarray}
\lefteqn{
16 \pi^2 \frac{d}{d \ln E}({\tilde D}) =
N_{\mbox{\scriptsize c}} {\Bigg \{}
2 \mbox{Tr} \left[ {\tilde D} {\tilde G}^{(D)\dagger}
+ {\tilde G}^{(U)} {\tilde U}^{\dagger} \right] {\tilde G}^ {(D)}
+ \mbox{Tr} \left[ {\tilde U} {\tilde U}^ {\dagger} +
{\tilde D} {\tilde D}^ {\dagger} \right] {\tilde D}
  }
\nonumber\\
&&
- \frac{1}{2} (\cot \beta) 
\mbox{Tr} \left[ {\tilde G}^{(D)} {\tilde D}^{\dagger}
- {\tilde D} {\tilde G}^{(D)\dagger} \right] {\tilde D}
+ \frac{1}{2} (\cot \beta) 
\mbox{Tr} \left[ {\tilde G}^{(U)} {\tilde U}^{\dagger}
- {\tilde U} {\tilde G}^{(U)\dagger} \right] {\tilde D} {\Bigg \}}
\nonumber\\
&& + {\Bigg \{} \frac{1}{2} \left[ {\tilde U} {\tilde U}^{\dagger}
+ {\tilde D} {\tilde D}^{\dagger} + {\tilde G}^{(U)} {\tilde G}^ {(U)\dagger}
+ {\tilde G}^{(D)} {\tilde G}^ {(D)\dagger} \right] {\tilde D} +
{\tilde D} \left[ {\tilde D}^{\dagger} {\tilde D} +
{\tilde G}^{(D)\dagger} {\tilde G}^{(D)} \right] 
\nonumber\\
&&
 - 2 {\tilde U} {\tilde U}^{\dagger} {\tilde D}
- 2 {\tilde G}^{(U)} {\tilde U}^ {\dagger} {\tilde G}^ {(D)} 
- A_D {\tilde D} {\Bigg \}} \ ,
\label{RGED}
\end{eqnarray}
\begin{eqnarray}
\lefteqn{
16 \pi^2 \frac{d}{d \ln E} \left( {\tilde G}^{(U)}\right) =
N_{\mbox{\scriptsize c}} {\Bigg \{}
 \mbox{Tr} \left[ {\tilde G}^{(U)} {\tilde G}^ {(U)\dagger} +
{\tilde G}^{(D)} {\tilde G}^ {(D)\dagger} \right] {\tilde G}^{(U)}
 }
\nonumber\\
&&
- \frac{1}{2} (\tan \beta) 
\mbox{Tr} \left[ {\tilde G}^{(U)} {\tilde U}^{\dagger}
- {\tilde U} {\tilde G}^{(U)\dagger} \right] {\tilde G}^{(U)}
+ \frac{1}{2} (\tan \beta) 
\mbox{Tr} \left[ {\tilde G}^{(D)} {\tilde D}^{\dagger}
- {\tilde D} {\tilde G}^{(D)\dagger} \right] {\tilde G}^{(U)} {\Bigg \}}
\nonumber\\
&& + {\Bigg \{} \frac{1}{2} \left[ {\tilde U} {\tilde U}^{\dagger}
+ {\tilde D} {\tilde D}^{\dagger} + {\tilde G}^{(U)} {\tilde G}^ {(U)\dagger}
+ {\tilde G}^{(D)} {\tilde G}^ {(D)\dagger} \right] {\tilde G}^{(U)} +
{\tilde G}^ {(U)} \left[ {\tilde U}^{\dagger} {\tilde U} +
{\tilde G}^{(U)\dagger} {\tilde G}^{(U)} \right] 
\nonumber\\
&&
 - 2 {\tilde D} {\tilde G}^{(D)\dagger} {\tilde U}
- 2 {\tilde G}^{(D)} {\tilde G}^ {(D)\dagger} {\tilde G}^ {(U)} 
- A_U {\tilde G}^{(U)} {\Bigg \}} \ ,
\label{RGEGU}
\end{eqnarray}
\begin{eqnarray}
\lefteqn{
16 \pi^2 \frac{d}{d \ln E} \left( {\tilde G}^{(D)} \right) =
N_{\mbox{\scriptsize c}} {\Bigg \{}
 \mbox{Tr} \left[ {\tilde G}^{(U)} {\tilde G}^ {(U)\dagger} +
{\tilde G}^{(D)} {\tilde G}^ {(D)\dagger} \right] {\tilde G}^{(D)}
  }
\nonumber\\
&&
- \frac{1}{2} (\tan \beta) 
\mbox{Tr} \left[ {\tilde G}^{(D)} {\tilde D}^{\dagger}
- {\tilde D} {\tilde G}^{(D)\dagger} \right] {\tilde G}^{(D)}
+ \frac{1}{2} (\tan \beta) 
\mbox{Tr} \left[ {\tilde G}^{(U)} {\tilde U}^{\dagger}
- {\tilde U} {\tilde G}^{(U)\dagger} \right] {\tilde G}^{(D)} {\Bigg \}}
\nonumber\\
&& + {\Bigg \{} \frac{1}{2} \left[ {\tilde U} {\tilde U}^{\dagger}
+ {\tilde D} {\tilde D}^{\dagger} + {\tilde G}^{(U)} {\tilde G}^ {(U)\dagger}
+ {\tilde G}^{(D)} {\tilde G}^ {(D)\dagger} \right] {\tilde G}^{(D)} +
{\tilde G}^ {(D)} \left[ {\tilde D}^{\dagger} {\tilde D} +
{\tilde G}^{(D)\dagger} {\tilde G}^{(D)} \right] 
\nonumber\\
&&
 - 2 {\tilde U} {\tilde G}^{(U)\dagger} {\tilde D}
- 2 {\tilde G}^{(U)} {\tilde G}^ {(U)\dagger} {\tilde G}^ {(D)} 
- A_D {\tilde G}^{(D)} {\Bigg \}} \ .
\label{RGEGD}
\end{eqnarray}

\section{A numerical example of evolution}

Here we present one simple but hopefully typical
example of the RGE evolution of parameters in the 2HDM(III)
framework. For simplicity, we assumed:
\begin{itemize} 
\item there is no CP violation -- 
all original four Yukawa matrices 
${\tilde U}^{(j)}$, ${\tilde D}^{(j)}$ are real, and the
VEV phase difference $\eta$ is zero;
\item
the masses and Yukawa parameters of the first
quark generation as well as that of the leptonic sector
are neglected (the quark Yukawa mass matrices
are therefore $2 \times 2$).
\end{itemize}

For the boundary conditions to the RGE's, at
the evolution energy $E=M_Z$,
we took the CSY ansatz (\ref{FCNCcon1})-(\ref{FCNCcon2}),
with $\xi^{(u)}_{ij}=1=\xi^{(d)}_{ij}$ for all $i,j=1,2$
[note: $i=1$ refers now to the second quark family ({\it c,s\/}),
and $i=2$ to the third family ({\it t,b\/})]. 
For the ($2 \times 2$) orthogonal CKM mixing matrix $V$ we took 
$V_{12}(M_Z) = 0.045 = -V_{21}(M_Z)$. The values of other
parameters at $E=M_Z$ were chosen:\\
$\tan \beta=1.0$; $v \equiv \sqrt{v_1^2+v_2^2}=246.22$ GeV;\\
$\alpha_3 = 0.118$, $\alpha_2=0.332$, $\alpha_1=0.101$; \\
$m_c=0.77$ GeV, $m_s=0.11$ GeV, $m_b=3.2$ GeV, and $m_t=171.5$ GeV.\\ 
The latter values of
quark masses correspond to: $m_c(m_c)\approx 1.3$ GeV,
$m_s(1 \mbox{GeV}) \approx 0.2$ GeV, $m_b(m_b) \approx 4.3$ GeV,
and $m_t^{\mbox{\scriptsize phys.}} \approx 174$ GeV
[$m_t(m_t)\approx 166$ GeV]. For ${\alpha}_3(E)$ we
used two-loop evolution formulas, with threshold effect
at $E \approx m_t^{\mbox{\scriptsize phys.}}$ taken into
account; for ${\alpha}_j(E)$ ($j=1,2$) we used one-loop
evolution formulas. 

The described simplified framework resulted in $18$ coupled
RGE's [for $18$ real parameters: $v^2$, $\tan \beta$,
${\tilde U}_{ij}$, ${\tilde D}_{ij}$, ${\tilde G}^{(U)}_{ij}$,
${\tilde G}^{(D)}_{ij}$], with the
mentioned boundary conditions at $E=M_Z$.
The system of RGE's was solved numerically, using
Runge-Kutta subroutines with adaptive stepsize control
(given in \cite{WHPressetal}). 
The numerical results were cross-checked in several ways,
including the following: FORTRAN programs for the RGE evolution
and for the biunitary transformations were constructed
independently by two of the authors (S.S.H. and G.C.),
and they yielded identical numerical results presented
in this Section.

The results for the FCNC Yukawa parameter ratios
$U_{ij}(E)/U_{ij}(M_Z)$ and $D_{ij}(E)/D_{ij}(M_Z)$
($i \not= j$) are depicted in Fig.~9. 
\begin{figure}[htb]
\mbox{}
\vskip10.cm\relax\noindent\hskip-0.8cm\relax
\includegraphics{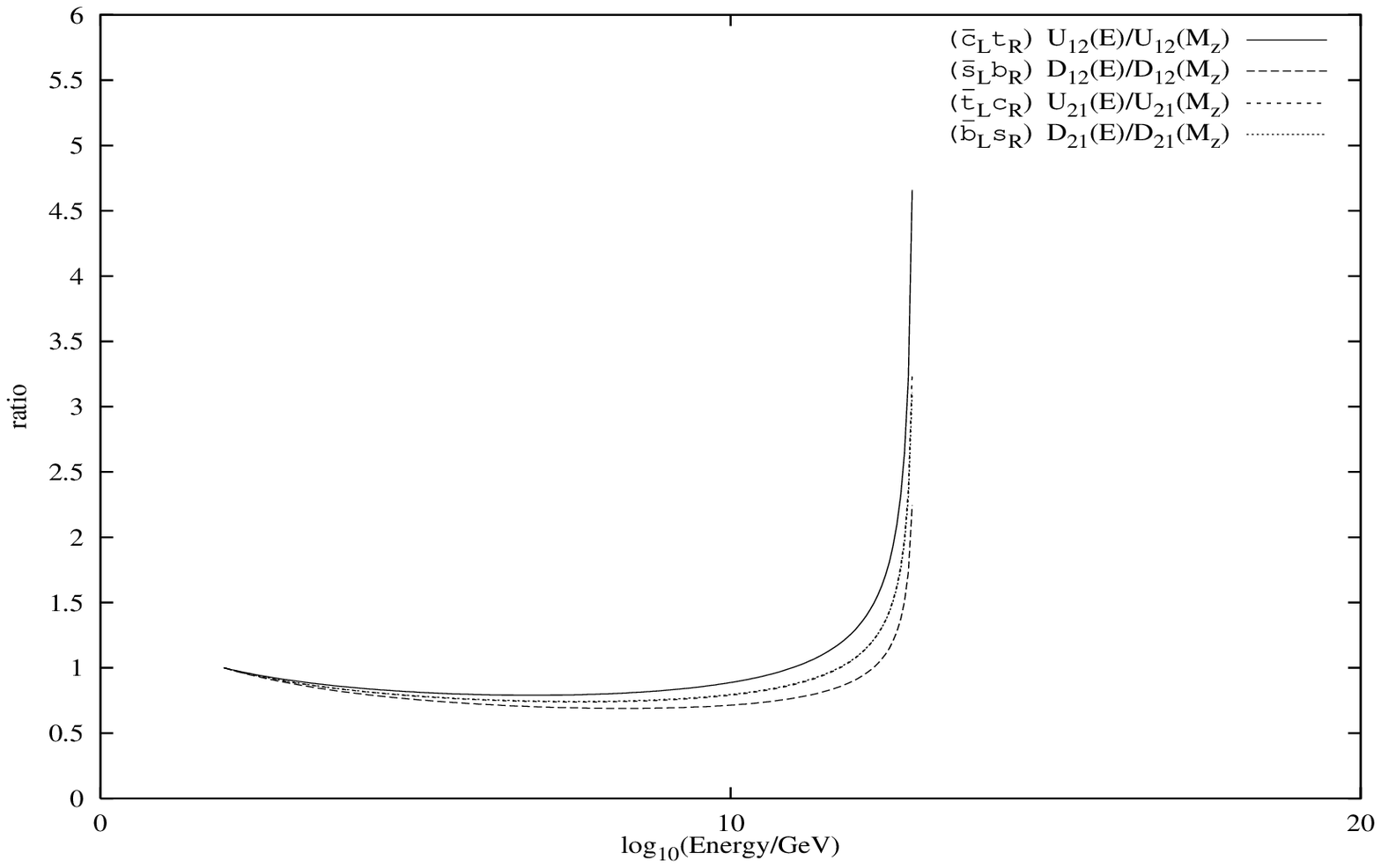} \vskip-0.4cm
\caption{\footnotesize Evolution of the
FCNC Yukawa parameter ratios $U_{ij}(E)/U_{ij}(M_Z)$,
$D_{ij}(E)/D_{ij}(M_Z)$ ($i \not= j$) in 2HDM(III). 
These parameters are in the quark mass basis. The choice of
parameters of the model at 
the starting low energy $E=M_Z$ is
specified in Sec.~4. Yukawa couplings of the first
generation were neglected; $i=1,2$ correspond to the
second and third quark generation, respectively.}
\end{figure}
{}From this
Figure we immediately notice that the FCNC coupling 
parameters of the down-type ($b$-$c$) sector
are remarkably stable as the evolution energy
increases. Even the up-type FCNC ratios, although
involving the heavy top quark, remain rather stable. 
Only very close to the top-quark-dominated
Landau pole ($E_{\mbox{\scriptsize pole}} \approx 
0.84 \cdot 10^{13}$ GeV)\footnote{
The value of $E_{\mbox{\scriptsize pole}}$
is strongly dependent on the given value of parameter 
$\xi$, as shown later in Fig.~13.}
the coupling parameters start to increase substantially.
For example, in the down-type FCNC sector ($b$-$c$)
the corresponding ratio $D_{21}(E)/D_{21}(M_Z)$
acquires its double initial value ({\it i.e.}, value $2$)
at $E \approx 0.7 E_{\mbox{\scriptsize pole}}$, which
is very near the (Landau) pole.\footnote{
Approximately the same is true also for
$U_{21}(E)/U_{21}(M_Z)$.} 
For the ratio $D_{12}(E)/D_{12}(M_Z)$
the corresponding energy is even closer to 
$E_{\mbox{\scriptsize pole}}$. 

\begin{figure}[htb]
\mbox{}
\vskip10.cm\relax\noindent\hskip-0.8cm\relax
\includegraphics{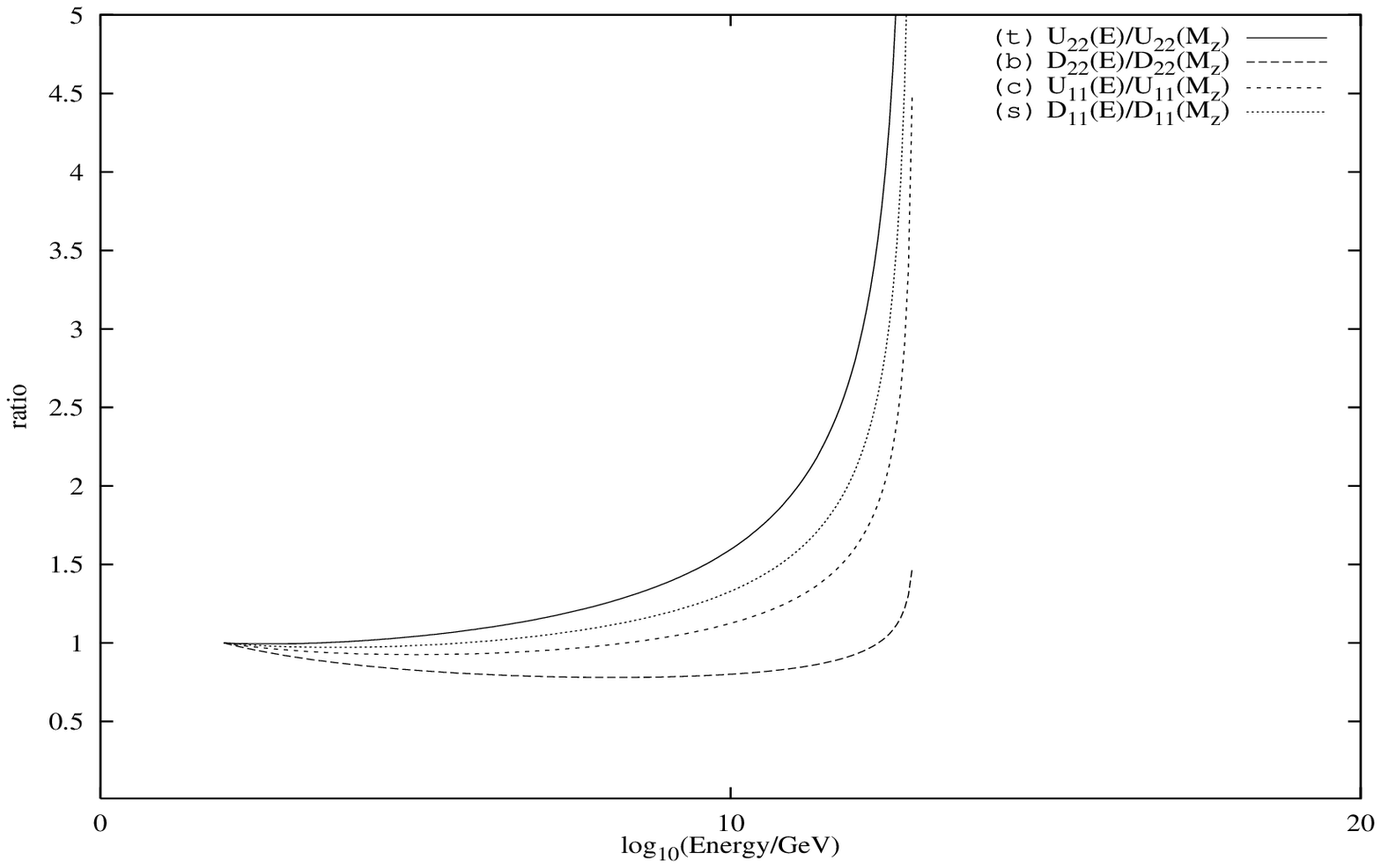} \vskip-0.4cm
\caption{\footnotesize Same as in Fig.~9,
but for the neutral current Yukawa coupling parameters 
$U_{jj}$ and $D_{jj}$ ($j=1,2$) which don't change flavor.}
\end{figure}
One may ask whether the mentioned stability features
in the evolution of {\it all} Yukawa coupling parameters in the
framework. This doesn't seem to be the case.
For example, in Figs.~10 and 11 we depicted,
with the described case of initial conditions, evolution
of the Yukawa coupling ratios connected with no FCNC's:
$U_{jj}(E)/U_{jj}(M_Z)$, $D_{jj}(E)/D_{jj}(M_Z)$,
$G^{(U)}_{jj}(E)/G^{(U)}_{jj}(M_Z)$ and 
$G^{(D)}_{jj}(E)/G^{(D)}_{jj}(M_Z)$ ($j=1,2$).
We see that evolution behavior of many of these ratios,
in stark contrast to the FCNC parameter case of Fig.~9, is
far from stable. For example, the ratio $D_{11}(E)/D_{11}(M_Z)$,
connected with $s$ quark,
acquires double its initial value approximately at the value 
$0.1 E_{\mbox{\scriptsize pole}}$; the ``mass'' Yukawa ratio
$G^{(D)}_{22}(E)/G^{(D)}_{22}(M_Z)$, corresponding to $b$ quark,
acquires half its initial value at
$10^{-5} E_{\mbox{\scriptsize pole}}$. These energies
are therefore substantially further away from the Landau pole
than the corresponding energies for the FCNC Yukawa
down-type sector (cf.~previous paragraph), 
even on the logarithmic scale.

In Fig.~12 we depicted evolution of the quark masses
for the discussed case. Note that they are not
simply proportional to the ``mass'' Yukawa
parameters $G^{(U)}_{jj}(E)$ and $G^{(D)}_{jj}(E)$, because the
VEV $v(E)$ also evolves with energy [cf.~Eq.~(\ref{RGEv2})]. 
Logarithmic scale was chosen for the masses in order
to include $m_t(E)$ in the figure.
{}From this figure we see that we have a rather strong variation of
$m_b(E)$  [and also of $m_t(E)$] continuously
when the evolution energy increases. For example,
$m_b(E)$ reaches half its initial value
[$m_b(E)=m_b(M_Z)/2$] at $E \approx 10^{-6} E_{\mbox{\scriptsize pole}}$,
which is very far away from $E_{\mbox{\scriptsize pole}}$
-- this again contrasts with the behavior of FCNC Yukawa
parameters of Fig.~9. 

The discussed numerical example of the 2HDM(III)
framework tells us that there definitely exist choices of 
reasonably suppressed ({\it i.e.}, phenomenologically acceptable)
FCNC Yukawa coupling parameters at low energies
such that these parameters remain largely unchanged (suppressed)
up to energy regions very close to the Landau pole.
On the other hand, this behavior doesn't feature
in the entire sector of the Yukawa coupling parameters.

It should be stressed that these results are independent of the
chosen value of the VEV ratio $\tan \beta$ at $E=M_Z$.
This is connected with our choice of the CSY boundary conditions
(\ref{FCNCcon1})-(\ref{FCNCcon2}) at $E=M_Z$ for the
Yukawa matrices in the quark mass basis 
(${\xi}_{ij}^{(u)} = {\xi}_{ij}^{(d)} = 1$)
and the reality of the chosen CKM matrix at $E=M_Z$.
These boundary conditions result in real and 
$\beta$-independent Yukawa matrices ${\tilde U}$,
${\tilde D}$, ${\tilde G}^{(U)}$, ${\tilde G}^{(D)}$
in a weak [$SU(2)_L$] basis\footnote{
We chose at $E=M_Z$ the following weak basis:
${\tilde U}=U$, ${\tilde G}^{(U)} = G^{(U)}$,
${\tilde D} = V D$, ${\tilde G}^{(D)} = V G^{(D)}$,
where $V$ is the CKM matrix (at $E=M_Z$). According to
relations (\ref{Gs})-(\ref{UDs}), the reality of the
Yukawa matrices ${\tilde U}$, ${\tilde D}$, ${\tilde G}^{(U)}$
and ${\tilde G}^{(D)}$ at the low energy $E=M_Z$ would follow, 
for example, from:
the requirement of no CP violation in the Yukawa sector
({\it i.e.}, the original Yukawa matrices ${\tilde U}^{(j)}$ and
${\tilde D}^{(j)}$ are all real) {\em together with\/} 
the requirement of no CP violation in the scalar sector 
({\it i.e.}, the VEV phase difference $\eta=0$) at that low energy.} 
at $E=M_Z$. The RGE's (\ref{RGEU})-(\ref{RGEGD}) then imply that
these matrices remain real and independent of $\beta$
at any evolution energy $E$, and that also their
counterparts $U$, $D$, $G^{(U)}$ and $G^{(D)}$
in the quark mass basis, as well as
the CKM matrix $V$, remain real and independent of $\beta$
at any energy $E$. 
\begin{figure}[htb]
\mbox{}
\vskip10.cm\relax\noindent\hskip-0.8cm\relax
\includegraphics{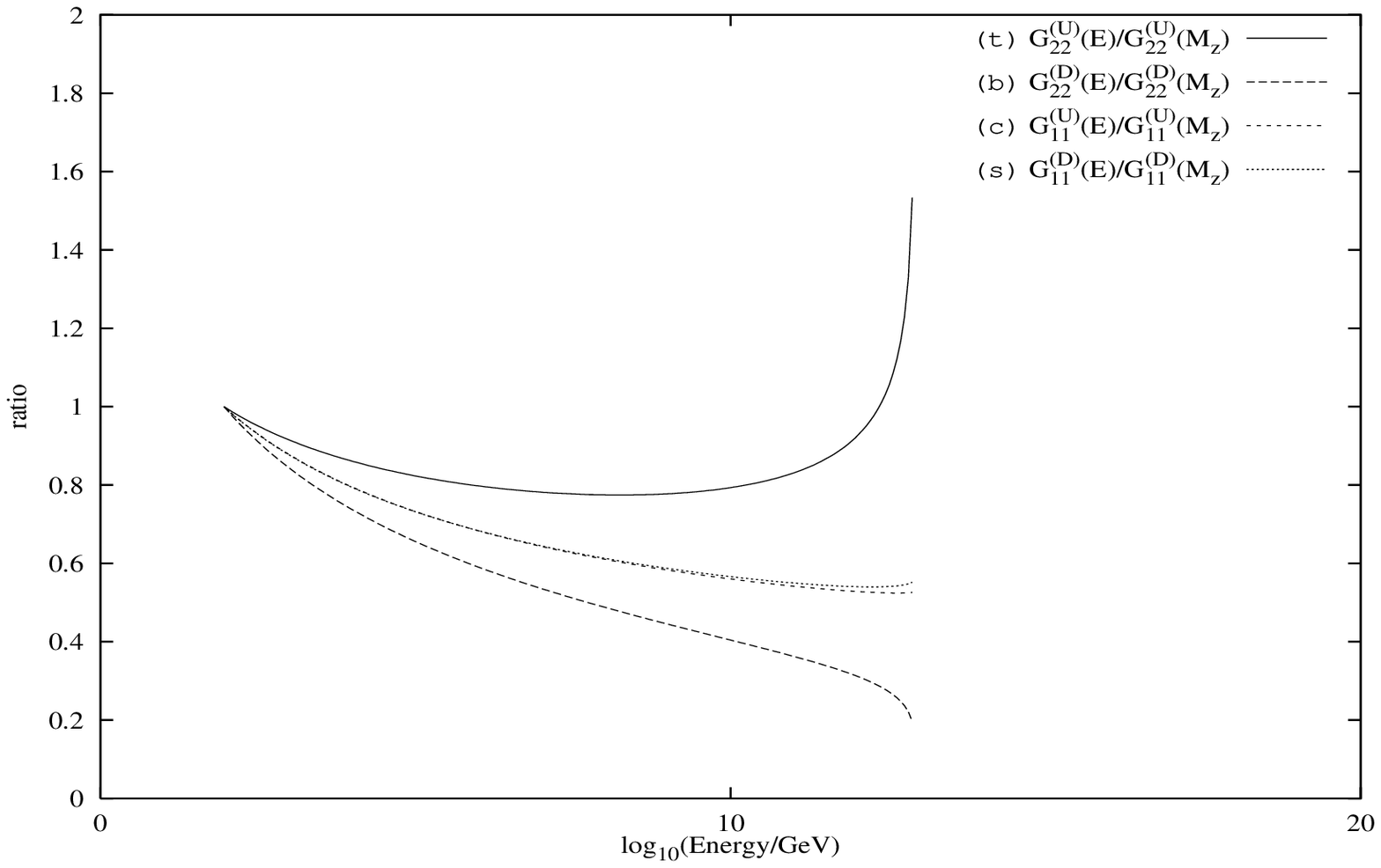} \vskip-0.4cm
\caption{\footnotesize Same as in Fig.~9,
but for the ``mass'' Yukawa parameters $G^{(U)}_{jj}$
and $G^{(D)}_{jj}$ ($j=1,2$) instead.
Since $G^{(U)}(E)$ and $G^{(D)}(E)$ matrices are diagonal
by definition (quark mass basis), these neutral current
Yukawa coupling parameters have zero FCNC components
automatically.}
\end{figure}
Stated otherwise, if there is
$\beta$-independence and no CP violation 
(neither in original Yukawa matrices
nor in the scalar sector) at a low energy ($E=M_Z$),
then these properties persist at all higher energies of 
evolution.\footnote{
CP conservation in the pure scalar sector at a low
energy $E=M_Z$ ({\it i.e.}, $\eta =0$) also
persists then at all higher energies of evolution,
since $d \eta/d \ln E = 0$ by the
reality of the Yukawa matrices, according to RGE
(\ref{RGExi}).}

This feature is in stark contrast with the situation in
the 2HDM(II) where the Yukawa matrices strongly depend on
$\beta$ already at low energies -- e.g., $g_t(M_Z)
= m_t(M_Z) \sqrt{2}/v_u = m_t(M_Z) \sqrt{2}/[v 
\sin (\beta(M_Z))]$. Also the location of the Landau pole
in the 2HDM(II) then crucially depends on $\beta(M_Z)$ --
smaller $\beta(M_Z)$ implies larger $g_t(M_Z)$ and hence
a drastically lower Landau pole.
\begin{figure}[htb]
\mbox{}
\vskip8.cm\relax\noindent\hskip-0.8cm\relax
\includegraphics{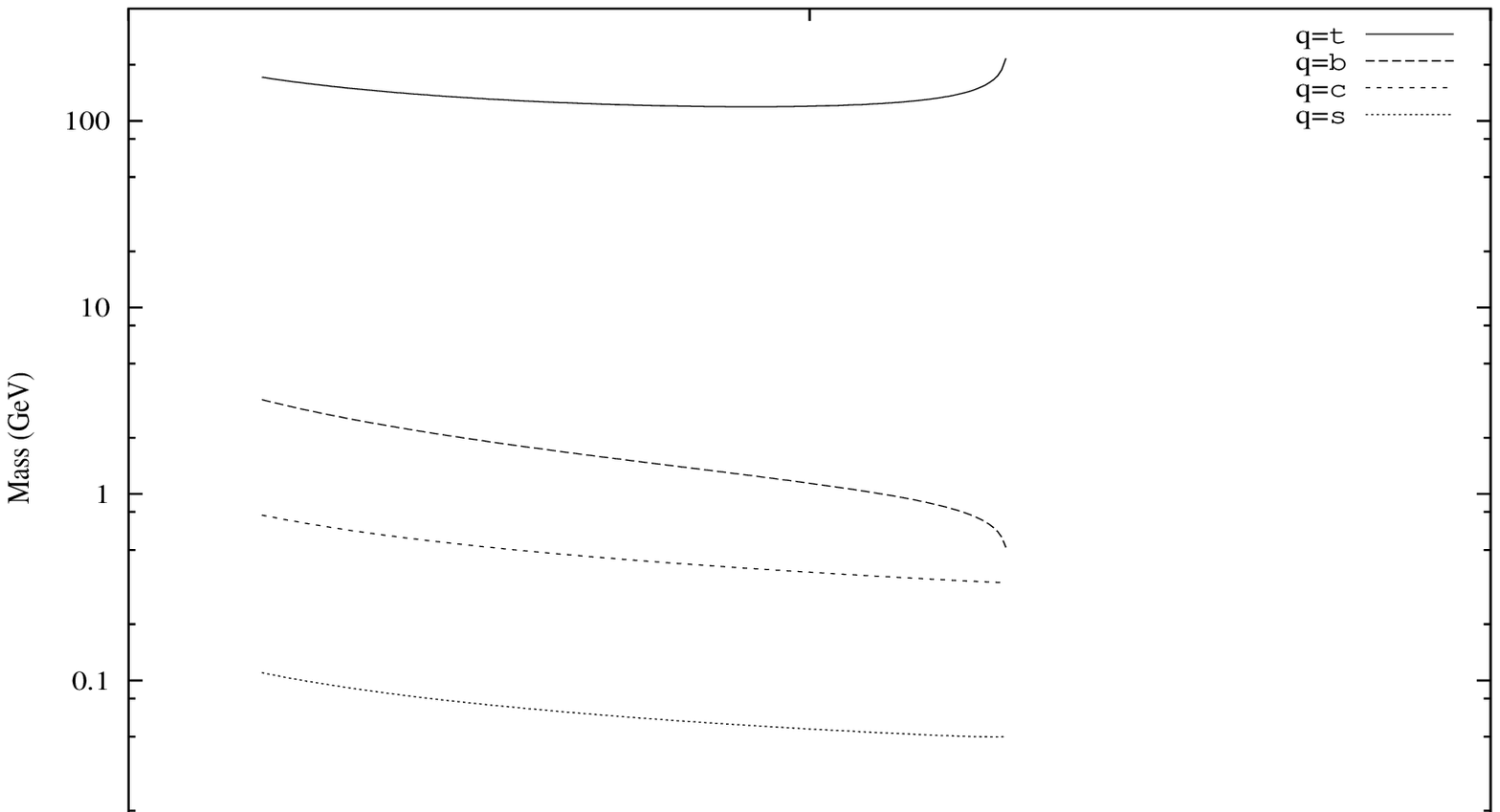} \vskip1.4cm
\caption{\footnotesize Evolution of the
quark masses $m_j^{(u)}(E) = G^{(U)}_{jj}(E) v(E)/\sqrt{2}$ 
and $m_j^{(d)}(E) = G^{(D)}_{jj}(E) v(E)/\sqrt{2}$
($m_1^{(d)}=m_s$, $m_2^{(d)}=m_b$; $m_1^{(u)}=m_c$,
$m_2^{(u)}=m_t$), for the discussed numerical example.}
\end{figure}

On the other hand, the 2HDM(III) framework
treats the up-type and the down-type sectors of quarks 
(the two VEV's $v_1$ and $v_2$) non-discriminatorily.
Therefore, it should be
expected that any reasonable boundary conditions for
Yukawa coupling parameters at low energies should also be independent
of $\beta$ in such frameworks, and this independence
then persists to a large degree
also at higher energies. Also the locations of the
Landau poles ({\it i.e.}, of the approximate scales of 
the onset of new physics)
should then be expected to be largely $\beta$-independent.
In this sense, 2HDM(III) has more similarity to the 
minimal SM (MSM) than to the 2HDM(II).
The persistence of complete $\beta$-independence of the
Yukawa coupling parameters at high energies
and of the Landau poles, however, can then be ``perturbed'' by
CP violation -- because
RGE's (\ref{RGEU})-(\ref{RGEGD}) are somewhat
$\beta$-dependent when the Yukawa matrices
$\tilde U$, {\it etc.}, are not real. Also the VEV
phase difference $\eta$ is then not a constant when
the energy increases [cf.~(\ref{RGExi})].

In addition to the connection between (low energy) CP violation
and $\beta$-dependence of
high energy results, there is yet another feature that
distinguishes the 2HDM(III) framework from the MSM
-- the Landau pole of a 2HDM(III) framework is in
general much lower than that of the MSM. We can see
that in the following way: let us consider that
only the Yukawa parameters connected with the
top quark degree of freedom are substantial,
{\it i.e.}, $G^{(U)}_{22} = g_t \sim 1$ and
$U_{22} = g_t^{\prime} \sim 1$. We have:
$g_t(E) = m_t(E) \sqrt{2}/v(E)$, as in the MSM, and
$g_t^{\prime}(E)$ is an additional large Yukawa parameter
-- both crucially influence location of the
Landau pole. Inspecting RGE's (\ref{RGEU}) and
(\ref{RGEGU}) for this special approximation
of two variables $g_t$ and $g_t^{\prime}$, we
see that RGE for $g_t$ is similar to that in the
MSM, but with an additional large positive
term on the right: $(3/2) (g_t^{\prime})^2 g_t$.
RGE for $g_t^{\prime}$ has a similar structure as
RGE for $g_t$, but with substantially larger coefficients
at the positive terms on the right. As a result,
$g_t^{\prime}(E)$ is in general larger than
$g_t(E)$. Our specific numerical example
[cf.~Figs.~10 and 11 for $U_{22}$ and $G^{(U)}_{22}$]
shows that $g_t^{\prime}(E)$ is on average
(average over the whole evolution energy range)
almost twice as large as $g_t(E)$. If we then
simply replace in the mentioned additional
term $(3/2) (g_t^{\prime})^2 g_t$
the parameter $(g_t^{\prime})^2$ by
$3.5 g_t^2$, we obtain from the resulting
``modified'' MSM RGE for $g_t$ a value for the Landau pole
in the region of $10^{12}-10^{13}$ GeV, which
is roughly in agreement with the actual value of the Landau
pole of our numerical example $E_{\mbox{\scriptsize pole}}
\approx 0.84 \cdot 10^{13}$ GeV. 
\begin{figure}[htb]
\mbox{}
\vskip10.cm\relax\noindent\hskip-0.8cm\relax
\includegraphics{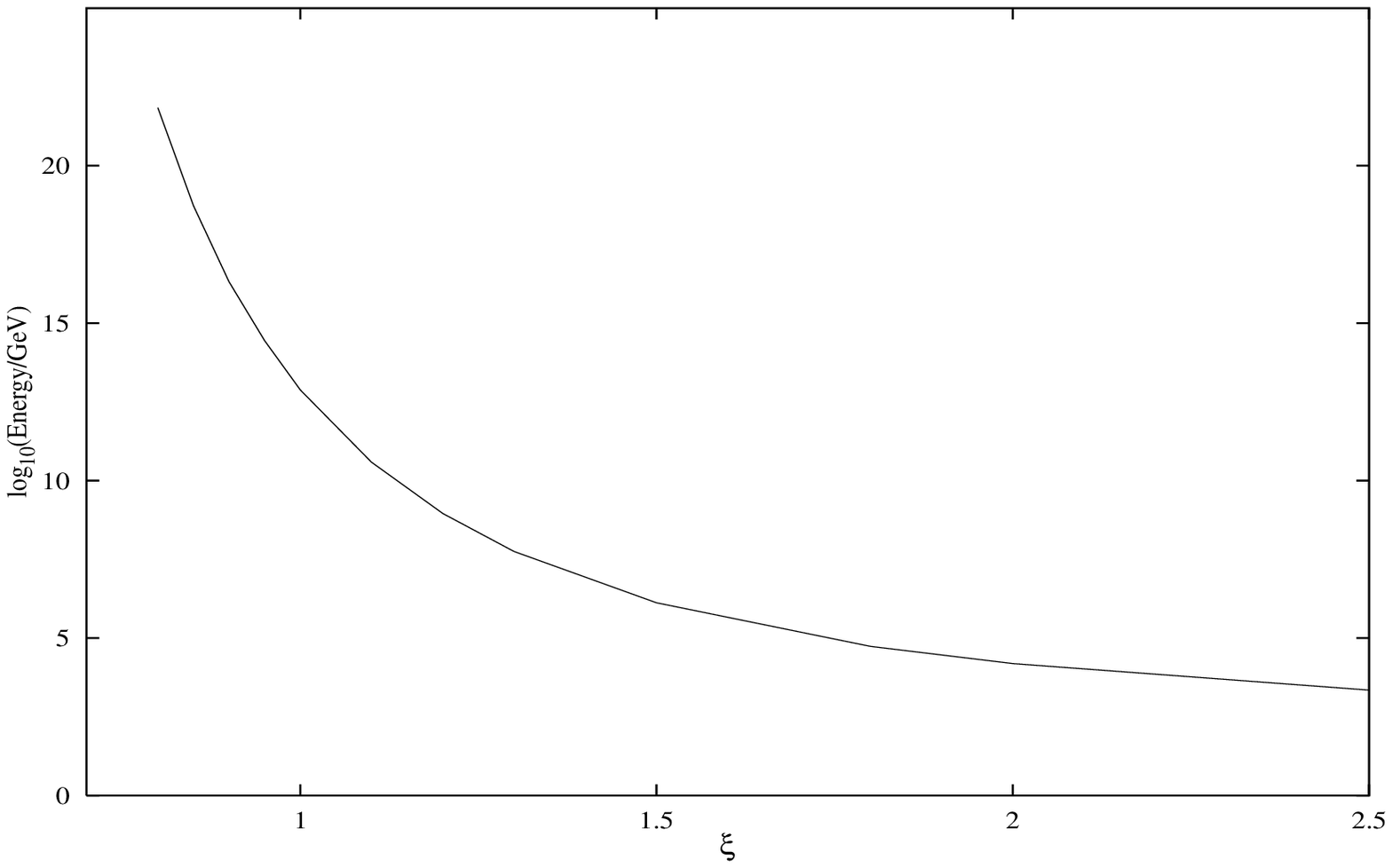} \vskip-0.4cm
\caption{\footnotesize Variation of the Landau pole energy
when the low energy ($E=M_Z$) parameters
${\xi}_{ij}^{(u)}={\xi}_{ij}^{(d)} \equiv \xi$ of the
CSY ansatz (\ref{FCNCcon1})-(\ref{FCNCcon2}) are
varied. For $\xi = 2.5$, the onset scale of new physics is already
quite low: $E_{\mbox{\scriptsize pole}} \approx 2$ TeV.}
\end{figure}
And this value is
much lower than $E_{\mbox{\scriptsize pole}}$
in the MSM which is above the Planck scale.
Of course, when we allow the ${\xi}_{ij}^{(u)}$
parameters of the CSY ansatz 
(\ref{FCNCcon1})-(\ref{FCNCcon2}) at $E=M_Z$
to deviate
from 1, we obtain larger $\log (E_{\mbox{\scriptsize pole}})$
for smaller ${\xi}_{ij}^{(u)}$, and smaller
$\log (E_{\mbox{\scriptsize pole}})$ for larger
${\xi}_{ij}^{(u)}$. In Fig.~13 we depicted this
variation of the Landau pole energy when the
CSY low energy parameters are varied.

\section{Summary and Conclusions}

We have derived the one-loop RGE's for the quark Yukawa
coupling matrices and the VEV's in the Standard Model framework
with the most general two-Higgs-doublet Yukawa
sector [2HDM(III)]. A simple -- and at low
energies phenomenologically acceptable --
numerical example for
the resulting evolution of these Yukawa parameters
suggests that the framework cannot be dismissed as
``unnatural'' from the FCNC-RGE point-of-view. Stated otherwise,
the numerical example shows remarkable stability of the
suppressed FCNC Yukawa coupling parameters 
when the energy of probes increases continuously
all the way to the vicinity of the top quark Landau pole.
The Landau pole is in general well below the Planck
scale in this framework.
We believe that further numerical 
investigations are warranted, in order to see whether
and/or to what degree this behavior survives when
we scan over certain reasonable 
(phenomenologically acceptable) ranges the values of
low energy parameters of the model --
{\it i.e.}, the Yukawa coupling parameters and $\tan \beta$
at $E \sim E_{\mbox{\scriptsize ew}}$.

\section{Abbreviations used in the article:}

{\noindent
AHR -- Antaramian, Hall and Ra\v sin;}

{\noindent
CKM -- Cabibbo, Kobayashi and Maskawa;}

{\noindent
CSY -- Cheng, Sher and Yuan;}

{\noindent
FCNC -- flavor-changing neutral current;}

{\noindent
LHS -- left-hand side;}

{\noindent
MSM -- minimal Standard Model;}

{\noindent
RHS -- right-hand side;}

{\noindent
RGE -- renormalization group equation;}

{\noindent
SM -- Standard Model;}

{\noindent
VEV -- vacuum expectation value;}

{\noindent
1PI -- one-particle-irreducible;}

{\noindent
1PR -- one-particle-reducible;}

{\noindent
2HDM -- two-Higgs-doublet (Standard) Model;}

{\noindent
2HDM(III) -- general two-Higgs-doublet (Standard) Model -- ``type III''.}

\vspace{0.5cm}

\section{Acknowledgments}
%The work of GC was supported in part by the Deutsche Forschungsgemeinschaft.
The work of GC was supported in part by the Bundesministerium fuer Bildung, 
Wissenschaft, Forschung und Technologie, Project. No. 057DO93P(7).
The work of CSK and SSH was supported 
%in part by the KOSEF, Project No. xxxxx,
%in part by Non-Directed-Research-Fund, KRF, Project No. xxxxx,
in part by the CTP of SNU, 
in part by Yonsei University Faculty Research Fund of 1997, 
in part by the BSRI Program, Ministry of Education, 
Project No. BSRI-97-2425, and 
in part by the KOSEF-DFG large collaboration
project, 
Project No. 96-0702-01-01-2.

\end{document}